\documentclass[pra,twocolumn,amsmath,amssymb,superscriptaddress]{revtex4-2}
\usepackage{epsfig,amsmath}
\usepackage{subfigure}
\usepackage{graphicx}
\usepackage{dcolumn}
\usepackage{stmaryrd}
\usepackage{mathrsfs}
\usepackage{pifont}
\usepackage{amsthm}
\usepackage{amssymb}
\usepackage{bm}
\usepackage{latexsym}
\usepackage{color}
\usepackage{epstopdf}
\usepackage{amsfonts}
\usepackage{units}
\usepackage{times}
\usepackage{upgreek}
\usepackage[colorlinks=true,linkcolor=blue,citecolor=blue,urlcolor=blue]{hyperref}
\begin{document}
	
\title{Dynamical generation of stable optical-microwave squeezing in structured reservoirs}
	
\author{Chen Wang}
\affiliation{College of Physics and Hebei Key Laboratory of Photophysics Research and Application, Hebei Normal University, Shijiazhuang, Hebei 050024, China}

\author{Man Shen}
\affiliation{College of Physics and Hebei Key Laboratory of Photophysics Research and Application, Hebei Normal University, Shijiazhuang, Hebei 050024, China}

\author{Shi-fan Qi}
\email{qishifan@hebtu.edu.cn}
\affiliation{College of Physics and Hebei Key Laboratory of Photophysics Research and Application, Hebei Normal University, Shijiazhuang, Hebei 050024, China}
	
\author{Yan-Kui Bai}
\email{ykbai@semi.ac.cn}
\affiliation{College of Physics and Hebei Key Laboratory of Photophysics Research and Application, Hebei Normal University, Shijiazhuang, Hebei 050024, China}
	
\date{\today}

\begin{abstract}
Two-mode squeezed states as paradigmatic entangled resources have broad applications in quantum information processing. Here, we study the generation of stable optical-microwave squeezing in structured environments within a hybrid electro-optomechanical system, where a mechanical oscillator is simultaneously coupled to an optical cavity mode and a microwave mode of an LC resonator. Specifically, an effective Hamiltonian that captures the optical-microwave squeezing interaction is constructed by combining strongly modulated driving fields applied to both photonic modes with a mechanical parametric amplifier. Based on this effective model, the dynamical evolution of two-mode squeezing in structured environments is analyzed. It is remarkably shown that the non-Markovian noise can substantially enhance the squeezing level in comparison to the Markovian case, and that two-mode squeezing can persist even in the absence of external driving fields under non-Markovian conditions, thereby mitigating the detrimental effects of anti-squeezing. Furthermore, the persistence of the two-mode squeezed state is enhanced when the environmental spectral densities of the microwave and optical modes are identical. Our work provides a theoretical framework for generating and persisting two-mode squeezing in structured environments. 
\end{abstract}
\maketitle

\section{Introduction}\label{intoduction}
Two-mode squeezed states (TMSSs) are prototypical entangled states in continuous-variable systems, playing a central role in quantum information processing~\cite{information,quantuminformation}, quantum communication~\cite{quantumnetworks,quacommunication}, and quantum metrology~\cite{quantum,metrology,metro}. Many protocols have been proposed to generate TMSS with high squeezing level (SL)~\cite{spinensembles,thermalenvironment,Squeezinglimit,twosqueezing,Quantum-enhanced,Quantumentanglement,87Rb,photon-phononsqueezing,antiferromagnetic,antiferromagneticmagnon}, with experimental realizations spanning thermal gases~\cite{long-livedentanglement}, ultracold atomic Bose–Einstein condensates~\cite{twin-atomstates,AtomicSpin,Single-Atom,SpinorBose}, atomic mechanical oscillators~\cite{AtomicMechanical}, spin ensembles in cavities~\cite{BosonicKitaevModels,BosonicPair,spinmodels}, and superconducting circuits~\cite{Superconducting,TravelingWave,WavePhonons}. In particular, by bridging two complementary regions of the electromagnetic spectrum, the optical-microwave TMSS~\cite{Entanglingmicrowaveswithlight} combines the long-distance transmission advantages of optical photons with the precise controllability of microwave modes~\cite{convert,Nonreciprocalconversion}, thereby underpinning scalable quantum networks~\cite{energyconversion,networks,computing}, secure quantum communication~\cite{securequantum}, and advanced quantum sensing technology~\cite{electro-opto-mechanical}. Therefore, the generation of optical-microwave TMSS with high SL remains a topic that requires further study.

Among various platforms, electro-optomechanical (EOM) systems~\cite{Bidirectionalconverter,EOMsystem,optomechanical,Transducer,optoelectromechanical} provide a versatile interface for generating optical-microwave TMSS by coupling optical and microwave modes via a mechanical intermediary~\cite{ReversibleOM,converter}. Two main approaches have been proposed within this framework. One employs reservoir engineering~\cite{opticssqueezed,QuantumInterference,Entanglingtwomicrowave} to dissipatively steer the system into a TMSS, while the other constructs an effective two-mode squeezing interaction~\cite{opcircuit}. In Markovian environments, since the phonon decay rate is typically much lower than that of the photon~\cite{Cavityopto,Coherentstatetransfer,EntanglingMechanicalMotion,Quantumsqueezing,OptomechanicalInteraction,CoherentCoupling,Activeoptomechanics}, the former yields limited SL, while the latter can generate high SL squeezing but with enhanced anti-squeezing~\cite{dynamicstable}, which hinders the precise measurement of TMSS. Both approaches are restricted to Markovian environments, raising the question of whether moving beyond the Markovian regime can overcome these limitations and enable high SL TMSS. Recently, non-Markovian processes have attracted considerable interest~\cite{non-Markovianregime,non-mar,Quantumnon-mar,non-Markoviandensity}, with non-Markovian spectral densities having been observed in micro-optomechanical systems~\cite{Non-Markovian}. Such environments can enhance quantum state conversion~\cite{statestransfer,nonStatetransfer,photon-phononconversion}, mechanical squeezing~\cite{non-Markoviansqueezing}, and quantum entanglement~\cite{Nonequilibrium,entanglement}, thereby motivating the exploration of non-Markovian memory effects for facilitating the generation of stable TMSS. In addition, previous works rely on external driving, while the persistence of TMSS after the driving is removed remains largely unexplored.
 
In this paper, we study the generation of stable optical-microwave TMSS in an EOM system under structured environments, described by an effective Hamiltonian that accounts for mechanically assisted two-mode squeezing via drive-enhanced optomechanical and electromechanical interactions. Using the non-Markovian Heisenberg-Langevin (NMHL) equation~\cite{CHENG2019385}, we analyze the dynamical process of optical-microwave squeezing in the non-Markovian environments. Our results show that the structured environments can generate dynamically stable TMSS and significantly enhance the SL compared to Markovian noise. Moreover, structured environments can sustain TMSS in the absence of external driving, effectively suppressing anti-squeezing and keeping it at a controlled level. Furthermore, we find that optimal persistence is achieved when the spectral densities of the optical and microwave modes are matched.

The rest of this work is organized as follows. In Sec.~\ref{model}, we introduce a three-mode EOM system and derive an effective Hamiltonian for optical-microwave squeezing assisted by a mechanical mode. Sec.~\ref{ppsqueeze} presents a theoretical framework for analyzing the squeezing dynamics in structured environments. Section~\ref{result} numerically investigates the generation and persistence of optical-microwave TMSS under non-Markovian conditions. Finally, Sec.~\ref{conclu} discusses experimental feasibility and summarizes the main results.

\section{EOM system and the effective Hamiltonian}\label{model}
\begin{figure}[t]
\centering
\includegraphics[width=0.75\linewidth]{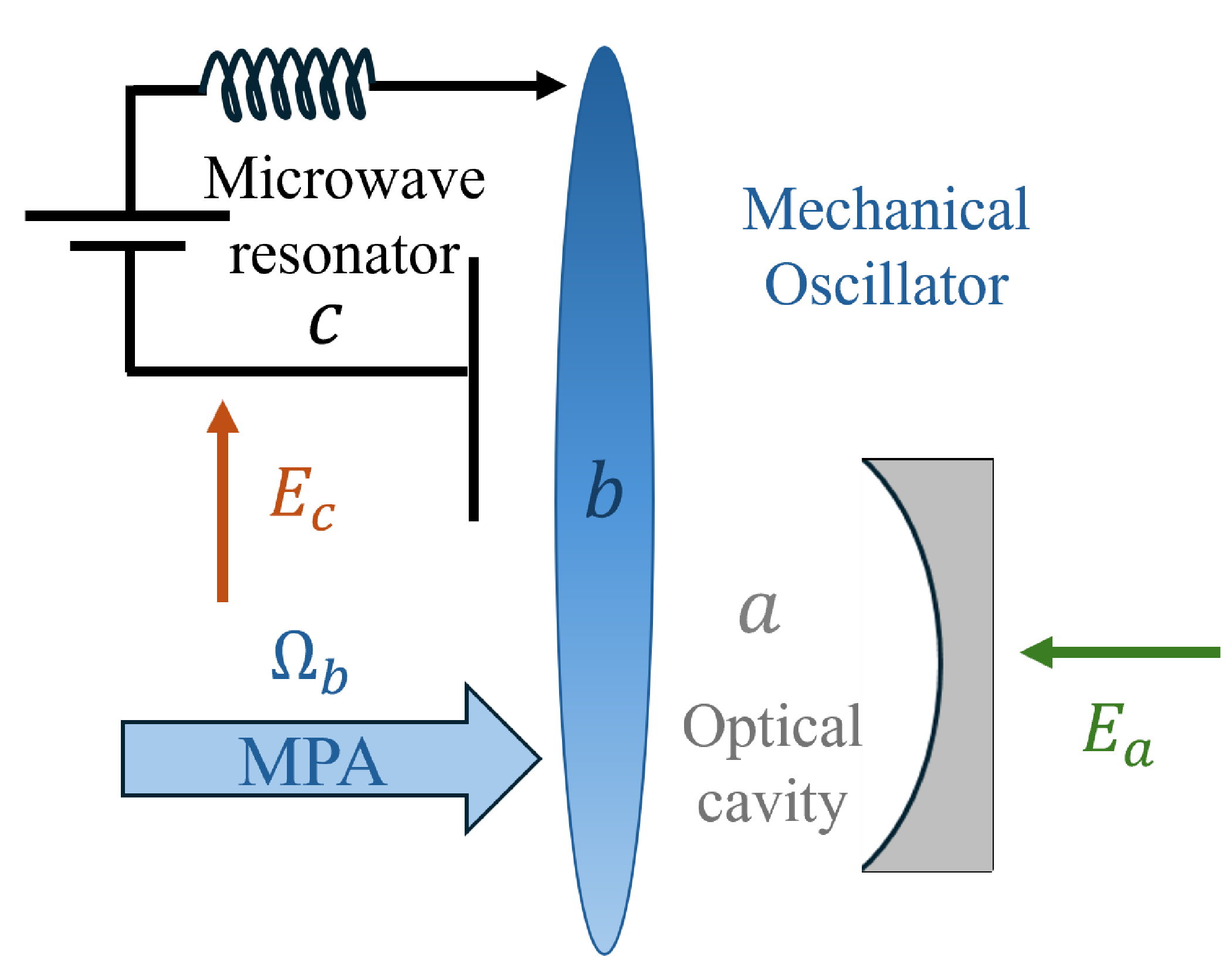}
\caption{Schematic diagram: The mechanical resonator serves as an intermediate mode $b$, coupling simultaneously to the microwave resonator $c$ and optical cavity $a$. The optical and microwave modes are driven by strong external fields $E_a$ and $E_c$, respectively, while the mechanical mode is subjected to a nonlinear driving with amplitude $\Omega_b$, giving rise to mechanical parametric amplification.}
\label{system}
\end{figure}
The schematic of the hybrid three-mode EOM system~\cite{Bidirectionalconverter,electro-opto-mechanical,EOMsystem,optoelectromechanical} under consideration is shown in Fig.~\ref{system}. The system comprises a microwave resonator, a single-mode optical cavity, and a mechanical oscillator. The mechanical mode couples to the optical cavity via radiation-pressure interaction and to the microwave mode of an LC circuit through capacitive coupling. Additionally, a nonlinear pump field is applied to the mechanical oscillator, leading to mechanical parametric amplification (MPA). Meanwhile, both the optical cavity and the microwave resonator are coherently driven by strong and proper classical external fields. The system Hamiltonian can be written as 
\begin{equation}\label{generalH}
\begin{aligned}
H_s&=\sum_{o=a,c}\omega_oo^\dag o+g_o o^\dag o(b+b^\dag)+iE_o(o^\dag e^{-i\varepsilon_ot}-oe^{i\varepsilon_ot})\\
&\quad+\omega_b b^\dag b+i\Omega_{b}(b^{\dag2} e^{-2i\varepsilon_bt}-b^2e^{2i\varepsilon_bt}),\\
\end{aligned}	
\end{equation}
where $a$ ($a^\dagger$), $b$ ($b^\dagger$), and $c$ ($c^\dagger$) denote the annihilation (creation) operators of the optical, mechanical, and microwave modes with transition frequencies $\omega_a$, $\omega_b$, and $\omega_c$, respectively. The parameters $g_a$ and $g_c$ are the single-excitation optomechanical and electromechanical coupling strengths. $E_a$ ($E_c$) and $\varepsilon_a$ ($\varepsilon_c$) are the Rabi and driving frequencies of the external drive applied to the optical (microwave) mode, respectively. The parameter $\Omega_b$ is the amplitude of the parametric drive on the mechanical mode, and $2\epsilon_b$ is the corresponding pump frequency.

Under the transformation $U(t)=\exp(it\sum_{s=a,b,c}\epsilon_ss^\dag s)$, the system Hamiltonian in Eq.~\eqref{generalH} transforms into
\begin{equation}
\begin{aligned}
H'_s&=\sum_{o=a,c}\Delta_oo^\dag o+g_o o^\dag o(b+b^\dag)+iE_o(o^\dag-o)\\
&\quad+\Delta_b b^\dag b+i\Omega_b(b^{\dag2}-b^2),
\end{aligned}
\end{equation}
where $\Delta_s\equiv\omega_s-\epsilon_s$, $s=a,b,c$. Under strong coherent drivings, the microwave and optical modes are assumed to acquire large steady-state amplitudes $|\langle a\rangle|\gg1$ and $|\langle c\rangle|\gg1$, respectively. This allows the system dynamics to be linearized by decomposing the operators $o=\langle o\rangle+\delta o, o=a,c$, where $\delta o$ represents the quantum fluctuation operator of mode $o$ around its mean value~\cite{Cavityopto}. In this case, the Hamiltonian $H'_s$ turns into
\begin{equation}\label{linHamini}
\begin{aligned}
H&=\Delta_a \delta a^\dag \delta a+\Delta_bb^\dag b+\Delta_c\delta c^\dag \delta c+i\Omega_b(b^{\dag2}-b^2)\\
&\quad+G(e^{-i\alpha}\delta a+e^{i\alpha}\delta a^\dag)(b+b^\dag)\\
&\quad+g(e^{-i\varphi}\delta c+e^{i\varphi }\delta c^\dag)(b+b^\dag),
\end{aligned}
\end{equation}
where $G=g_a|\langle a\rangle|$ and $g=g_c|\langle c\rangle|$ are the driving-enhanced optomechanical and microwave-mechanical coupling strengths, respectively. The parameters $\alpha$ and $\varphi$ represent the corresponding phases, defined by $e^{i\alpha}=\langle a\rangle/|\langle a\rangle|$ and $e^{i\varphi}=\langle c\rangle/|\langle c\rangle|$. In deriving Eq.~\eqref{linHamini}, all linear terms are eliminated by appropriately tuning the detunings, since they correspond to average displacements of the modes. For simplicity, we henceforth adopt the convention $\delta a\to a$ and $\delta c\to c$.

In the rotating frame with the unitary transformation
$S=\exp[ir(b^2+b^{\dag 2})/2]$, the quadratic terms about $b^2$ and $b^{\dag 2}$ in Eq.~\eqref{linHamini} are eliminated by choosing $\tanh(2r)=2\Omega_b/\Delta_b$. Under this condition, the Hamiltonian in Eq.~\eqref{linHamini} reduces into
\begin{equation}\label{linHam}
\begin{aligned}
H&=H_0+V,\\
H_0&=\Delta_aa^\dag a+\tilde{\omega}_bb^\dag b+\Delta_cc^\dag c,\\
V&=Ge^r(e^{-i\alpha}a+e^{i\alpha}a^\dag)(b+b^\dag)\\
&+ge^r(e^{-i\varphi }c+e^{i\varphi }c^\dag)(b+b^\dag),
\end{aligned}
\end{equation}
where $\tilde{\omega}_b=\Delta_b/\cosh(2r)$. The parameter $r$ represents the effect of the MPA.

In the large detuning regime, i.e., $ge^r,Ge^r\ll\{|\tilde{\omega}_b-\Delta_a|,|\tilde{\omega}_b-\Delta_{c}|\}$, and under the near-opposite condition $\Delta_a=-\Delta_c+\delta$ with $\delta$ being the energy shift, an effective Hamiltonian describing the optical-microwave squeezing can be obtained using perturbation theory~\cite{photon-phononconversion,PhotonAtoms,atomsphotons}. The resulting effective Hamiltonian is
\begin{equation}\label{effHam}
\begin{aligned}
H_{\rm eff} = g_{\rm eff}[e^{-i(\alpha+\varphi)}ac+e^{i(\alpha+\varphi)}a^\dag c^\dag],
\end{aligned}
\end{equation}
where the effective coupling strength $g_{\rm eff}$ is
\begin{equation}\label{effcoustr}
\begin{aligned}
g_{\rm eff}=\frac{2\tilde{\omega}_bgGe^{2r}}{\Delta_c^2-\tilde{\omega}_b^2}.
\end{aligned}
\end{equation}
The details are presented in Appendix~\ref{appeffective}. 

We assess the validity of the effective model by diagonalizing the full system transition matrix and comparing the analytical expressions for $g_{\rm eff}$ and $\delta$ in Eq.~\eqref{effcoustr} with numerical simulations (see Appendix~\ref{validity}). This analysis identifies an approximate parameter regime in which the effective Hamiltonian provides a valid and reliable description of the system dynamics and yields a relatively large effective coupling, namely, $0.1\tilde{\omega}_b\le g, G\le 0.3\tilde{\omega}_b$ and $r\le 0.2$. Within this regime, the effective coupling strength satisfies $g_{\rm eff} \ge 0.01\tilde{\omega}_b$. Moreover, consistent with the theoretical predictions of Eq.~\eqref{effcoustr}, increasing the coupling strengths $g$ and $G$, as well as the MPA parameter $r$, can further enhance the magnitude of $g_{\rm eff}$.

\section{Optical-microwave squeezing in structured environments}\label{ppsqueeze} 
In this section, we analyze the influence of structured environments on the squeezing process by employing the NMHL equation, assuming the system interacts with an environment modeled as a collection of independent harmonic oscillators. The total Hamiltonian is given by
\begin{equation}\label{toth}
H_{\rm tot}=H_{\rm eff}+H_E+H_I,
\end{equation}
where $H_{\rm eff}$ is the effective Hamiltonian given in Eq.~\eqref{effHam}. The environmental Hamiltonian $H_E$ corresponding to the optical-microwave system are
\begin{equation}\label{envh}
H_E=\sum_{k}\omega_ka_k^\dag a_k+\sum_{j}\tilde{\omega}_jc_j^\dag c_j,
\end{equation}
where $\omega_k$ and $\tilde{\omega}_j$ denote the transition frequencies of the $k$-th optical and the $j$-th microwave reservoir modes, respectively. The two reservoirs are assumed to be uncorrelated. The interaction Hamiltonian between the system and the environment is described by
\begin{equation}\label{inth}
H_I=\sum_kg_ka_k^\dag ae^{-i\omega_at}+\sum_jJ_jc_j^\dag c e^{-i\omega_c t}+{\rm H.c}.,
\end{equation}
where $g_k$ and $J_j$ are the system-reservoir coupling strengths for the optical and microwave modes, respectively. It is written under the rotating wave approximation, which is valid when the coupling strengths are much smaller than the transition frequencies, i.e., $g_k,J_j\ll\omega_a,\omega_c$.

With the effective Hamiltonian in Eq.~\eqref{effHam} and the interaction Hamiltonian in Eq.~\eqref{inth}, one can derive the NMHL equation
\begin{equation}\label{nmhl}
\dot{O}(t)=TO(t)-\int_0^t\bar{F}(t-s)O(s)\mathrm{d}s+\epsilon_{in}(t).
\end{equation}
Here, the time-evolution operator for the system modes is $O(t)=[a(t),c^\dag(t)]^T$ and the input noise operator is $\epsilon_{in}(t)=[a_{in}(t),c^\dag_{in}(t)]^T$, where $a_{in}\equiv-i\sum_kg_ke^{-i (\omega_k-\omega_a)t}a_k(0)$
and $c^\dag_{in}\equiv i\sum_jJ_je^{i(\tilde{\omega}_j-\omega_c)t}c^\dagger_j(0)$. The initial input noise operators $a_{in}(0)$ and $c_{in}(0)$ satisfy $\langle a_k^\dagger(0)a_{k'}(0)\rangle=\bar{n}_a(\omega_k)\delta_{kk'}$ and $\langle c_j^\dagger(0) c_{j'}(0)\rangle=\bar{n}_c(\tilde{\omega}_j)\delta_{jj'}$, respectively, where $\bar{n}_a(\omega_k)=1/[\exp(\hbar\omega_k/k_BT_a)-1]$ and $\bar{n}_c(\tilde{\omega}_j)=1/[\exp(\hbar\tilde{\omega}_j/k_BT_c)-1]$ are the thermal average occupation numbers of the corresponding reservoir modes. The drift matrices $T$ and $\bar{F}(t)$ are derived as
\begin{equation}\label{TFmatrices}
\begin{aligned}
T&=ig_{\text{eff}}\begin{bmatrix} 
	0& -e^{i(\alpha+\varphi)}\\e^{-i(\alpha+\varphi)}&0
\end{bmatrix},\\
\bar{F}(t)&=\begin{bmatrix} 
	f_a(t) & 0 \\
	0 & f_c^*(t) 
\end{bmatrix},
\end{aligned}
\end{equation}
respectively, where $f_a(t)=\sum_kg_k^2 e^{-i(\omega_k-\omega_a)t}$ and $f_c(t)=\sum_jJ_j^2e^{-i(\tilde{\omega}_j-\omega_c)t}$ are the time correlation functions. By introducing the spectral density functions, the functions can be written as 
\begin{equation}\label{spectfunc}
f_o(t)=\int d\omega J_o(\omega)e^{-i(\omega-\omega_o)t},\quad o=a,c,
\end{equation}
where $J_o(\omega)$ is the spectral density of mode $o$. For the microwave and optical modes, the spectral density is given by a Lorentzian form,
\begin{equation}\label{spectrum} J_o(\omega)=\frac{\gamma_o\lambda^2_o}{(\omega-\omega_o)^2+\lambda^2_o}, 
\end{equation} 
in which $\lambda_o$ is the width of the spectrum and $\gamma_o$ is the global dissipation rate~\cite{lorentz1,lorentzprotect,Newton}. Here, the central frequency of the environment associated with mode $o$ is assumed to coincide with the corresponding system's transition frequency $\omega_o$. The parameter $1/\lambda_o$ denotes the characteristic memory time of the environment.

For the Markovian case, the corresponding decoherence rate is $\kappa_o=\pi J_o(\omega_o)=\pi\gamma_o, o=a,c$, according to the the Weisskopf-Wigner theory~\cite{WWtheory}. In the presence of Markovian noise, it can be shown that an optimal two-mode squeezing operator exists for unsteady system dynamics,
\begin{equation}\label{MOSX}
	X=\cos\phi X_a+\sin\phi Y_c,\\
\end{equation}
where $\tan(2\phi)=2g_{\rm eff}/(\kappa_a-\kappa_c)$, and 
\begin{equation}\label{quadoperators}
	X_a=\frac{e^{-i\alpha}a+e^{i\alpha}a^\dag}{\sqrt{2}},\quad  Y_c=\frac{e^{-i\varphi}c-e^{i\varphi}c^\dag}{i\sqrt{2}},
\end{equation}
denote the corresponding orthogonal operators of optical and microwave modes, respectively. A detailed derivation is provided in Appendix~\ref{squeemarkovian}. The anti-two-mode squeezing operator, which is orthogonal to the squeezing operator $X$ in Eq.~\eqref{MOSX}, can be explicitly expressed as
\begin{equation}\label{MOSY}
	Y=-\sin\phi X_a+\cos\phi Y_c,
\end{equation}
which is utilized to assess the time-dependent behavior of the anti-two-mode squeezing effect in this scheme.

Subsequently, the variances of the quadrature operators $X$ in Eq.~\eqref{MOSX} and $Y$ in Eq.~\eqref{MOSY} under structured environments can be obtained by solving the evolution equation given in Eq.~\eqref{nmhl}. Formally, Eq.~\eqref{nmhl} can be solved by assuming $O(t)=\mathcal{U}(t)O(0)+\mathcal{V}(t)$, where $\mathcal{U}(t)$ is a 2×2 coefficient matrix, $\mathcal{U}(t)=[\mathcal{U}_{11}(t),\mathcal{U}_{12}(t);\mathcal{U}_{21}(t),\mathcal{U}_{22}(t)]$, and $\mathcal{V}(t)=[\mathcal{V}_1(t),\mathcal{V}_2(t)]^T$ is a vector of operators associated with the system's non-equilibrium Green's functions. Hence, the variances of  $X$ and $Y$  can be derived as
\begin{equation}\label{variXY}
\begin{aligned}
\Delta X&=\cos^2\phi V_{11}+\sin^2\phi V_{44}+\sin(2\phi)V_{14},\\
\Delta Y&=\sin^2\phi V_{11}+\cos^2\phi V_{44}-\sin(2\phi)V_{14},
\end{aligned}
\end{equation}
where the values of $V_{11}$, $V_{44}$, and $V_{14}$ are given in Eq.~\eqref{CMelement}, Appendix~\ref{appderivation}. The corresponding SL in the decibel unit is defined by
\begin{equation}\label{SL}
	S=-10\rm{log}\left(\frac{\Delta X}{\Delta X_{zp}}\right),
\end{equation}
where $\Delta X_{\rm zp}=0.5$ is the standard fluctuation in the zero-point level. This parameter is used for quantifying two-mode squeezing effects.

Since the system's evolution under structured environment is different from the evolution in Markovian noise, the optimal squeezing operator is no longer given by the theoretical prediction $X$ in Eq.~\eqref{MOSX}. The genuine optimal operator $\tilde{X}$ is obtained by numerically minimizing the general operator $\cos\tilde{\phi}X_a+\sin\tilde{\phi}Y_c$ with respect to the angle $\tilde{\phi}$. It is found that the variance $\Delta\tilde{X}$ of the optimal operator $\tilde{X}$ corresponds to the minimal eigenvalue of CM $V$ in Eq.~\eqref{CMelement}, which is given by
\begin{equation}
\Delta\tilde{X}=\frac{V_{11}+V_{44}-\sqrt{(V_{11}-V_{44})^2+4V_{14}^2}}{2},
\end{equation} 
and the corresponding maximal anti-squeezing variance is
\begin{equation}
\Delta\tilde{Y}=\frac{V_{11}+V_{44}+\sqrt{(V_{11}-V_{44})^2+4V_{14}^2}}{2}.
\end{equation} 
The SL associated with the genuine optimal operator $\tilde{X}$ in the decibel unit is defined by 
\begin{equation}\label{SL2}
	\tilde{S}=-10\rm{log}\left(\frac{\Delta\tilde{X}}{\Delta X_{\rm zp}}\right).
\end{equation}
Notice the variance $\Delta\tilde{X}=\Delta X$ for the Markovian case.

Furthermore, we analyze the persistence process of optical-microwave TMSS within the framework of open quantum systems, focusing on their evolution under environmental noise after the external driving fields are switched off. Once the external drive is removed, the effective coupling strength in the Hamiltonian vanishes ($g_{\rm eff}=0$), and the interaction between the optical (microwave) mode and the mechanical resonator becomes negligible. Consequently, each mode evolves almost independently, i.e., undergoing nearly free evolution.

\section{Numerical Results}\label{result} 
\subsection{Numerical Results for the Generation Process}
\begin{figure}[t]
	\centering
	\includegraphics[width=0.98\linewidth]{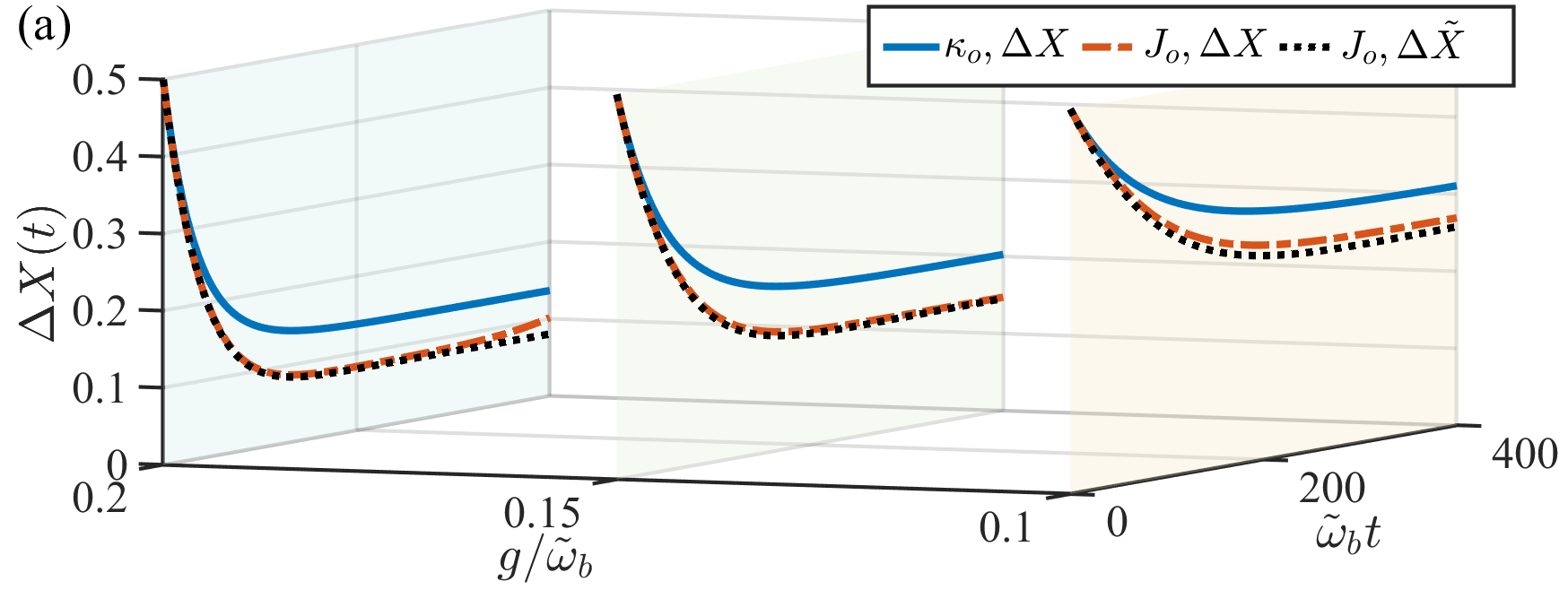}
	\includegraphics[width=0.48\linewidth]{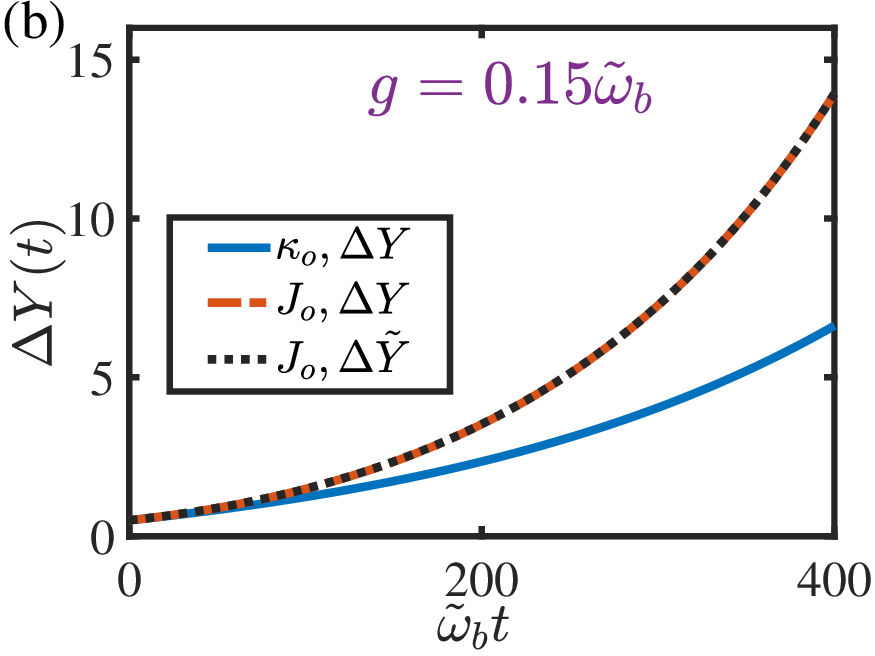}
	\includegraphics[width=0.48\linewidth]{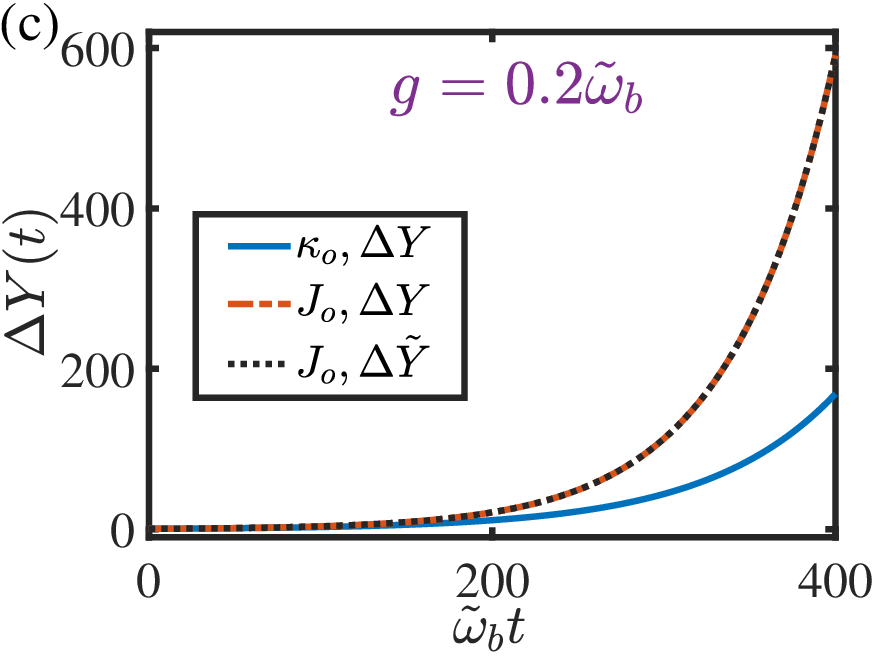}
	\caption{(a) Time evolution of the squeezing variance $\Delta X(t)$ for different environments and coupling strengths $g$. [(b), (c)] Time evolution of the anti-squeezing variance $\Delta Y(t)$ at $g=0.15\tilde{\omega}_b$ and $g=0.2\tilde{\omega}_b$ for different environments. The Markovian case is characterized by $\kappa_o$, whereas the structured environment is labeled by $J_o$. Here we fix $\Delta_c=3.5\tilde{\omega}_b$, $\alpha=\varphi=0$, $g=G$, and $r=0.2$ for all figures. The parameters are chosen as $\gamma_c=1.5\gamma_a$ with $\gamma_a=10^{-3}\tilde{\omega}_b$ and $\lambda_c=1.5\lambda_a$ with $\lambda_a=10^{-2}\tilde{\omega}_b$, while the thermal occupation numbers are $\bar{n}_a=\bar{n}_c=0$.}
	\label{MOXYgenerate}
\end{figure}
\begin{figure}[htbp]
	\centering
	\includegraphics[width=0.48\linewidth]{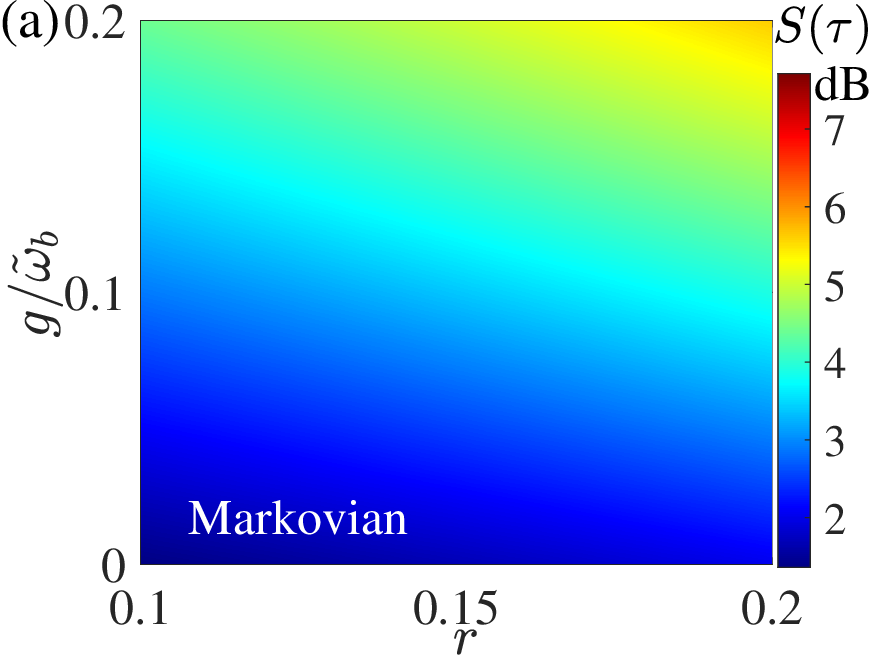}
	\includegraphics[width=0.48\linewidth]{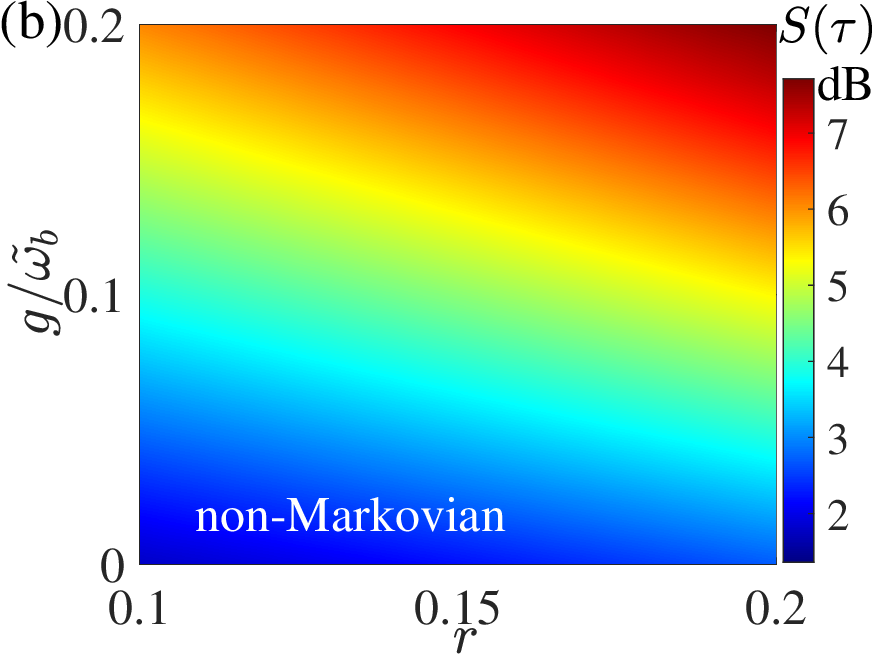}
	\includegraphics[width=0.48\linewidth]{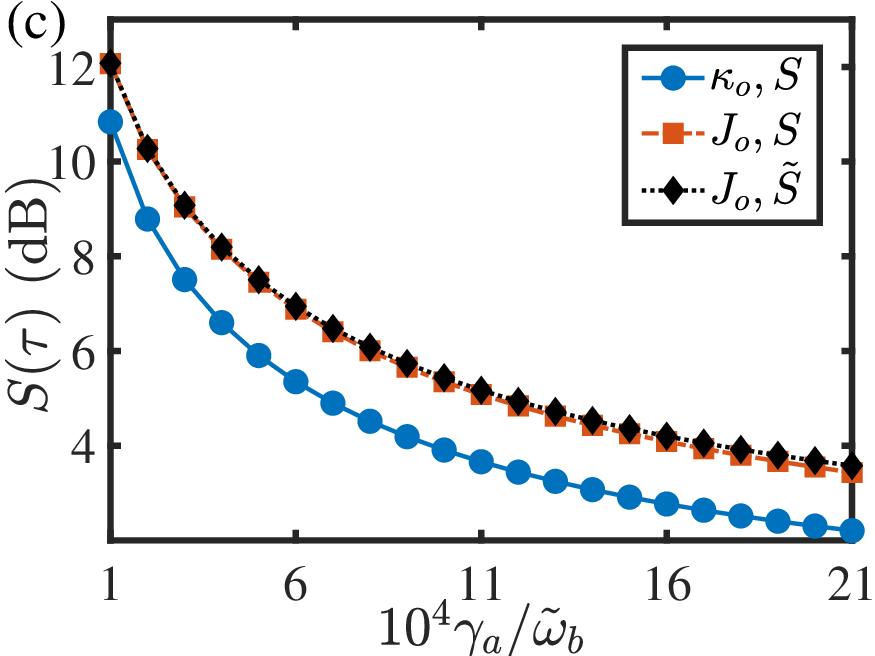}
	\includegraphics[width=0.48\linewidth]{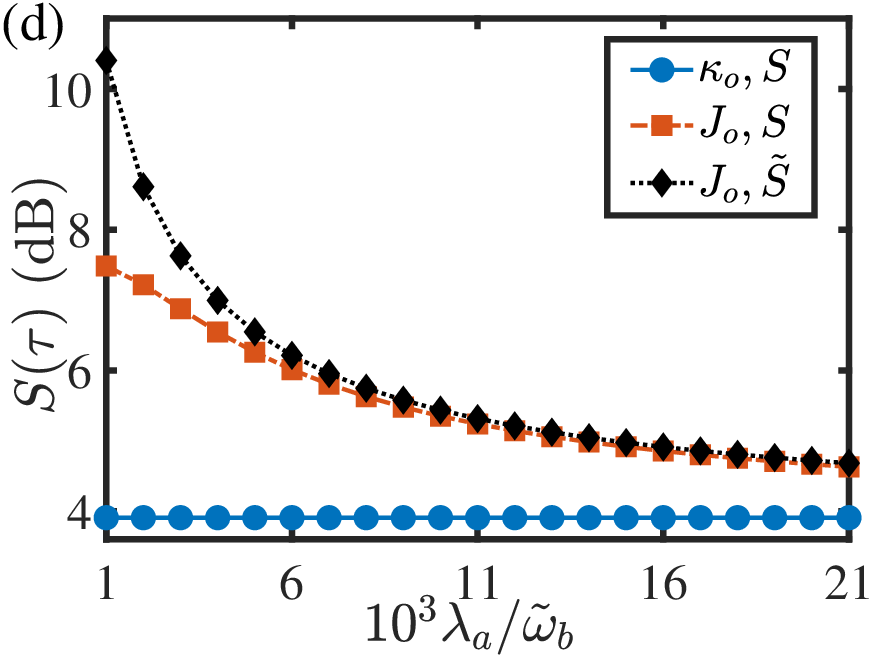}
	\caption{[(a), (b)] The squeezing level (SL) $S(\tau)$ at $\tilde{\omega}_b\tau=300$ is plotted in the parameter space spanned by the coupling strength $g$ and the MPA parameter $r$ for both Markovian and structured environments, respectively. [(c),(d)] The SL $S(\tau)$ is shown as a function of the global dissipation rate $\gamma_a$ and spectrum width $\lambda_a$ for various coupling strengths, respectively. In panels (c) and (d), we take $g=G=0.15\tilde{\omega}_b$ and $r=0.2$. The other parameters are identical to those in Fig.~\ref{MOXYgenerate}.}
	\label{MOXYgenerate2}
\end{figure} 

We begin by comparing the effect of environmental noise on the variance of the squeezing operator $X$ in Eq.~\eqref{MOSX} under both Markovian and structured environments, as shown in Fig.~\ref{MOXYgenerate}(a). It is observed that the variance $\Delta X$ gradually converges to a constant value over time, regardless of whether the system is embedded in a Markovian (blue solid line) or a structured environment (red dash-dotted line). In contrast, the variance $\Delta Y$ exhibits exponential growth, indicating unsteady system dynamics [see Figs.~\ref{MOXYgenerate}(b) and (c)]. These results demonstrate that asymptotic stable two-mode squeezing can persist even under unsteady evolution, whatever the type of environmental noise. Compared with the Markovian case, the structured environment yields a significantly higher squeezing, albeit at the expense of a concomitant increase in the anti-squeezing component. Moreover, one can observe that the genuine optimal variance $\Delta \tilde{X}$ (black dotted lines) matches well with the corresponding $\Delta X$ for $g=0.1\tilde{\omega}_b$ and $g=0.15\tilde{\omega}_b$, respectively, indicating that the operator $X$ in Eq.~\eqref{MOSX} accurately captures the effective two-mode squeezing phenomenon. For a larger coupling strength $g=0.2\tilde{\omega}_b$, $\Delta X$ shows a slight deviation from the optimal $\Delta \tilde{X}$ at long times. As the anti-squeezing effect becomes stronger for larger $g$, even a small deviation in the squeezing operator can lead to a pronounced discrepancy. Nevertheless, the operator $X$ in Eq.~\eqref{MOSX} remains an effective description of the optical-microwave squeezing over a relatively long time interval.

To characterize the protocol performance over a broad parameter range, we evaluate the SL $S(\tau)$ at the representative time $\tilde{\omega}_b\tau=300$, as shown in Figs.~\ref{MOXYgenerate2}(a) and (b). In both Markovian and structured environments, the SL $S(\tau)$ increases monotonically with the coupling strength $g$ and the MPA parameter $r$. Notably, the structured environment consistently yields a higher SL than the Markovian case. Specifically, the maximal SL in Markovian noises is $5.63$ dB at $g=0.2\tilde{\omega}_b$ and $r=0.2$, whereas it is enhanced to $7.71$ dB under non-Markovian noise. Moreover, for a fixed $r$, the enhancement becomes slightly more pronounced as $g$ increases, whereas for a fixed $g$, the improvement remains nearly constant across the considered range of $r$. 

In Fig.~\ref{MOXYgenerate2}(c), the SL $S(\tau)$ is plotted versus the global dissipation rate $\gamma_a$. Across all values of $\gamma_a$, the optical-microwave squeezing consistently achieves a higher SL in structured environments compared to Markovian cases, and the SL $S(\tau)$ in structured environments is close to the corresponding optimal value $\tilde{S}$. In addition, the enhancement of SL induced by the structured environment becomes more pronounced as $\gamma_a$ increases. Similarly, the SL under different spectrum widths $\lambda_a$ are depicted in Fig.~\ref{MOXYgenerate2}(d). In the Markovian case, the decay rate $\kappa_o$ is independent of the spectral width of the Lorentzian environment, resulting in a constant SL over the entire range of $\lambda_a$. By contrast, narrower spectral widths within structured environments, corresponding to stronger non-Markovian effects, lead to enhanced two-mode squeezing. However, they also result in a larger discrepancy between the SL $S$ and the optimal value $\tilde{S}$, making the squeezing operator $X$ given in Eq.~\eqref{MOSX} (obtained in Markovian environments) unable to fully capture the two-mode squeezing phenomenon. As the spectral width increases, the SL in structured environments gradually approaches the value in Markovian environments. This behavior is consistent with the theoretical understanding that the Markovian limit corresponds to an environment with an infinitely broad spectrum.

\subsection{Numerical Results for the persistence Process}
\begin{figure}[t]
	\centering
	\includegraphics[width=0.98\linewidth]{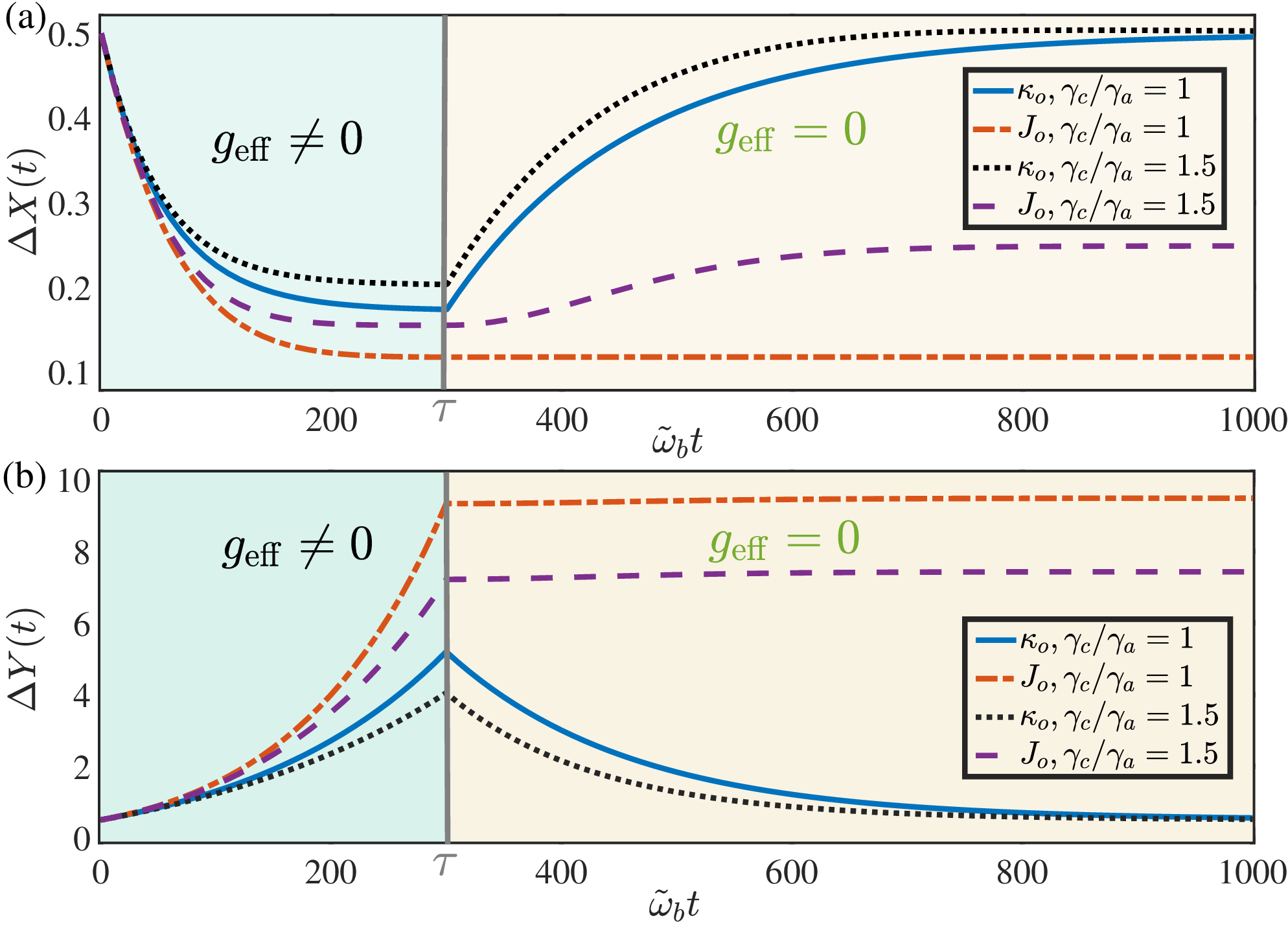}
	\caption{[(a), (b)] The time evolutions of the variances $\Delta X(t)$ and $\Delta Y(t)$ are shown for both the generation (teal areas) and persistence (beige areas) processes under different environmental noises, respectively. Here, the spectral-density parameters satisfy $\lambda_c/\lambda_a=\gamma_c/\gamma_a$, and all other parameters are identical to those in Figs.~\ref{MOXYgenerate}(b), except for $\gamma_c$ and $\lambda_c$.}
	\label{MOXY}
\end{figure}
Moreover, as shown in Fig.~\ref{MOXYgenerate}, while $\Delta X$ approaches a steady value, the variance $\Delta Y$ grows exponentially with time. This behavior primarily arises from the continuous pumping of the external driving fields, which keeps the system in an unsteady state. Such exponential growth poses significant challenges for precise experimental measurements and also leads to inefficient use of resources. To overcome these limitations, we investigate the sustainability process of optical-microwave TMSS after the external drives are switched off within the framework of open quantum systems. 

\begin{figure}[t]
	\centering
	\includegraphics[width=0.48\linewidth]{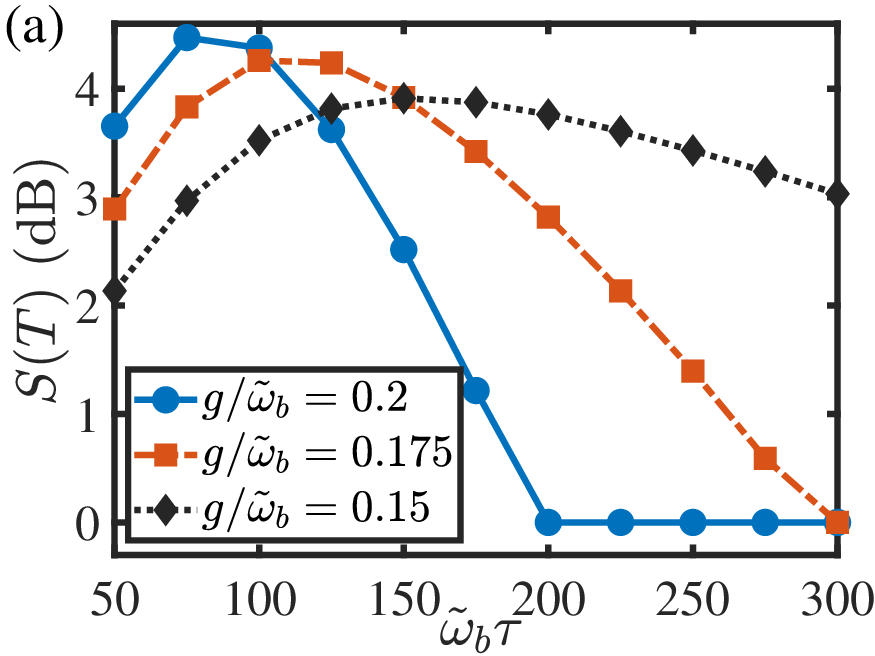}
	\includegraphics[width=0.48\linewidth]{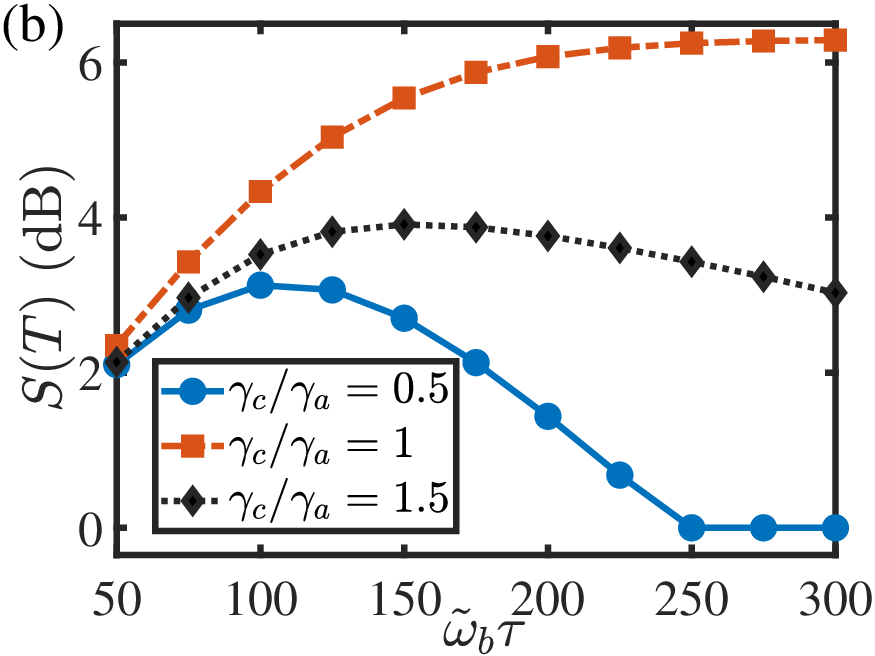}
	\includegraphics[width=0.48\linewidth]{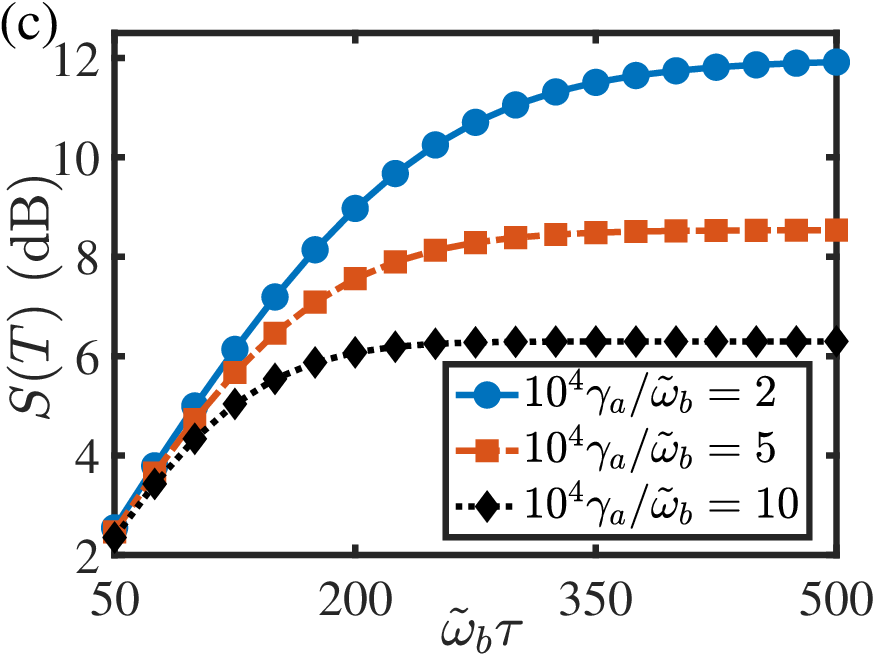}
	\includegraphics[width=0.48\linewidth]{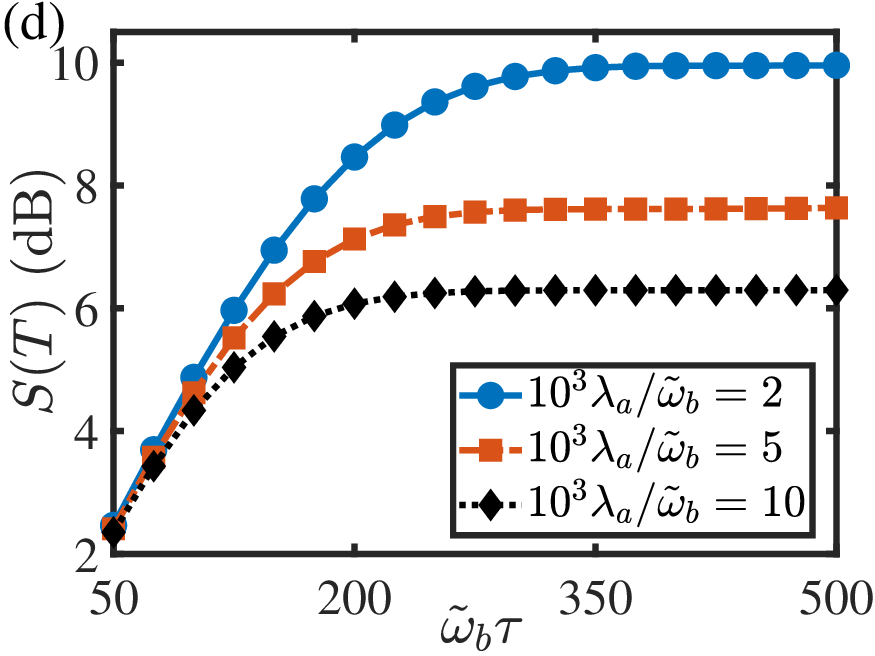}
	\caption{The squeezing level $S(T)$ at a uniformly chosen prolonged duration $T$ (with $\tilde{\omega_b}T=1000$) is plotted as a function of the driving-switch-off time $\tau$. In panel (a), $g=G$, and all other parameters are identical to those in Fig.~\ref{MOXYgenerate}(b). In panel (b), $g=G=0.15\tilde{\omega}_b$, $\gamma_a=10^{-3}\tilde{\omega}_b$, $\lambda_a=10^{-2}\tilde{\omega}_b$, and the ratio $\lambda_c/\lambda_a=\gamma_c/\gamma_a$. In panels (c) and (d), $g=G=0.15\tilde{\omega}_b$ and $\lambda_c/\lambda_a=\gamma_c/\gamma_a=1$, while all other parameters remain unchanged.}\label{SLpersistence}
\end{figure}

In Figs.~\ref{MOXY}(a) and (b), we present the variance dynamics of the squeezing operator $X$ and the anti-squeezing operator $Y$ over two distinct time intervals for different environmental spectral densities. The curves labeled $\gamma_c=\gamma_a$ correspond to identical environmental noise acting on modes $c$ and $a$. The first interval, $t\in[0,\tau]$ ($\tilde{\omega}_b\tau=300$), corresponds to the generation of optical-microwave squeezing (teal regions), during which the effective two-mode squeezing Hamiltonian in Eq.~\eqref{effHam} is established and the system evolves as in Figs.~\ref{MOXYgenerate}(a) and (b). At $t=\tau$, the driving fields are switched off, thereby eliminating the optical-microwave coupling, and the subsequent dynamics characterize the persistence of the generated squeezing (beige regions). Under Markovian noise, both variances $\Delta X$ [blue solid and black dotted lines in Fig.~\ref{MOXY}(a)] and $\Delta Y$ [blue solid and black dotted lines in Fig.~\ref{MOXY}(b)] relax to the common value $0.5$, corresponding to the vacuum variance $\Delta X_{\rm zp}$, indicating decay to the vacuum state. By contrast, in structured environments, memory effects induce a backflow of information from the environment to the system, leading to long-lived optical-microwave squeezing, as illustrated by the red dashed and purple dotted lines in Fig.~\ref{MOXY}(a). Furthermore, when the spectral densities of the two modes are matched [red dashed line in Fig.~\ref{MOXY}(a)], the variance $\Delta X(t)$ remains constant, indicating perfectly persistent two-mode squeezing.

In Fig.~\ref{SLpersistence}(a), we present the SL $S(T)$ at $\tilde{\omega}_bT=1000$ as a function of the driving switch-off time $\tau$, considering distinct structured environmental spectra for the microwave and optical modes. An optimal switch-off window is observed, within which the SL of the stored TMSS remains relatively high. As the coupling strength $g$ increases, this optimal window becomes narrower and shifts to earlier times. Nevertheless, a larger $g$ also leads to a higher achievable SL. In Fig.~\ref{SLpersistence}(b), we plot the SL $S(T)$ as a function of $\tau$ for different environmental spectra. Here, the optical spectrum $J_a$ is fixed, while the microwave spectrum $J_c$ is varied. The results show that the maximum SL is obtained when the spectra are identical, i.e., $J_a=J_c$ (red dashed line with squares). Under this condition, $S(T)$ increases with $\tau$ and eventually approaches a steady-state value. By contrast, when the spectra are mismatched, an optimal time window emerges for achieving higher squeezing level. However, even within this window, the attainable SL remains lower than that obtained under matched spectral conditions. Moreover, reducing the microwave dissipation rate (black dotted line with diamonds) further degrades the persistence performance of the TMSS. In Figs.~\ref{SLpersistence}(c) and (d), the SL $S(T)$ is plotted as a function of $\tau$ for different values of $\gamma$ and $\lambda$, assuming identical environmental spectra for both modes. It is evident that $S(T)$ increases monotonically with increasing $\gamma$ and $\lambda$. Moreover, $S(T)$ grows with $\tau$ and gradually approaches a stable value for larger $\gamma$ and $\lambda$.

We now provide a physical explanation for the persistence of TMSS in structured environments. Owing to the memory effect, the non-Markovian noise feeds back into the system. When the spectra of the two modes are identical, this feedback is symmetric, enabling coherent retrieval of the two-mode squeezing information stored in the environment and thereby preserving the squeezing. In contrast, when the two environmental spectra differ, the feedback becomes asymmetric, introducing anti-squeezing components that degrade the stored squeezing. This leads to a competition between squeezing and anti-squeezing effects. Although longer preparation times enhance the squeezing, they also amplify the anti-squeezing, resulting in an optimal switch-off time window for maximal persistence efficiency. Notably, increasing the coupling strength $g$ or reducing the dissipation intensity further strengthens the anti-squeezing effect, causing the optimal window to shift earlier and become narrower.

\section{Discussion and Conclusion}\label{conclu}
Our protocol is primarily based on an EOM system, and the parameters used in our analysis are within reach of current experimental implementations~\cite {Bidirectionalconverter,electro-opto-mechanical,EOMsystem,optoelectromechanical}. Specifically, the mechanical transition frequency is typically in the range $\omega_b/2\pi\sim 10-100$ MHz, with a decay rate $\kappa_b\sim 10^{-6}\omega_b$. Under strong driving fields, the enhanced microwave-mechanical and optomechanical coupling strengths are $g, G\sim0.1\omega_b$. Nonlinear driving applied to the mechanical oscillator (referred to as MPA) has also been successfully implemented in recent works~\cite{MPApara}. Additionally, both the microwave and optical thermal occupations are effectively negligible at low temperatures $T\sim 10$ mK, while the mechanical mode $N_b\sim 10$. For the structured environments considered in our study, the Lorentzian spectrum density of the environment has been widely used in the research on optical or microwave systems~\cite{lorentz1,lorentz1s,memory}. Moreover, beyond the EOM system, our scheme is extendable to other hybrid systems involving three bosonic modes. For example, photon-phonon squeezing in structured environments can be realized via a Kerr magnon interface~\cite{photon-phononconversion,photon-phononsqueezing}.

In summary, we have proposed a protocol for generating mechanical-mode-assisted microwave-optical squeezing within an EOM system, where a mechanical resonator is simultaneously coupled to a microwave mode and an optical cavity mode. This protocol is based on an effective two-mode squeezing Hamiltonian mediated by the mechanical mode. In the open-quantum-system framework, we study the dynamical evolution of optical-microwave TMSS generation governed by the effective Hamiltonian, using the NMHL equation. Our results demonstrate that a stable TMSS can be achieved under both Markovian and non-Markovian environmental noise, even beyond the conventional system stability regime, and that structured environments significantly enhance the SL. Moreover, we investigate the sustainability dynamics of the TMSS in the absence of external quantum control drivings. Remarkably, it is revealed that the memory effect associated with non-Markovian noise can sustain genuinely stable and high-quality optical-microwave squeezing without requiring additional control driving fields. Furthermore, our analysis demonstrates that spectral compatibility between the optical and microwave reservoirs is essential for achieving high-performance persistence of TMSS. Overall, our work provides an important implementation of TMSS generation in a hybrid EOM system under structured environments.

\section{Acknowledgments} This work was supported by National Science Foundation of China (Grants No. 12404405, No. 12404330, and No. 11575051), Hebei National Science Foundation (Grant No. A2021205020 and No. A2025205030), Hebei 333 Talent Project (No. B20231005), and the funds of Hebei Normal University (Grants No. L2024B10 and No. L2026J02).

\section{Data availability}
The data that support the findings of this article are not publicly available. The data are available from the authors upon reasonable request.

\appendix
\section{Derivation of the effective Hamiltonian} \label{appeffective}
This appendix presents the derivation of the effective Hamiltonian in Eq.~\eqref{effHam} based on near-degenerate perturbation theory~\cite{photon-phononconversion,PhotonAtoms,atomsphotons}. We focus on the parameter regime in which the optical and microwave detunings satisfy $\Delta_a\approx-\Delta_c$ and are both far detuned from the effective mechanical frequency $\tilde{\omega}_{b}$, such that $ge^r,Ge^r\ll|\tilde{\omega}_{b}-\Delta_a|,|\tilde{\omega}_{b}-\Delta_{c}|$. In this regime, the tensor-product Fock state $|nlk\rangle\equiv|{n\rangle}_a|{l\rangle}_b|{k\rangle}_c$ is nearly degenerate with $|(n+1)l(k+1)\rangle$. Here, the subscripts $a, b,c$ respectively represent the optical photon, mechanical oscillator, and microwave modes, respectively, while $n$, $l$, and $k$ label the corresponding Fock-state occupations. To second order in the perturbative coupling strengths $g$ and $G$, the effective coupling strength or energy shift between any two near-degenerate eigenstates of the unperturbed Hamiltonian $H_0$ in Eq.~\eqref{linHam} is given by
\begin{equation}\label{effcoustrequ}
\begin{aligned}
	\tilde{g}=\sum_{n\neq m,j}\frac{V_{jn}V_{nm}}{\omega_m-\omega_n},
\end{aligned}
\end{equation}
where $V_{nm}\equiv\langle n|V|m\rangle$ represents the matrix element of the interaction Hamiltonian $V$ between the unperturbed eigenstates $|n\rangle$ and $|m\rangle$, and $\omega_{n}$ is the corresponding eigenenergy of the state $|n\rangle$, with $V$ is considered as a perturbation to $H_0$. 

By applying the above second-order perturbation theory, one can obtain an analytical approximation to the effective Hamiltonian governing transitions between arbitrary basis states $|nlk\rangle$ and $|(n+1)l(k+1)\rangle$. This effective Hamiltonian takes the form
\begin{equation}\label{effHamequ}
\begin{aligned}
	H_\text{eff}&=\epsilon_1|nlk\rangle\langle nlk|\\
	&+(\Delta_a+\Delta_c+\epsilon_2)|(n+1)l (k+1)\rangle \langle (n+1)l(k+1)|\\
	&+\left[\tilde{G}|nlk\rangle\langle(n+1)l(k+1)|+\text{H.c.}\right].
\end{aligned}
\end{equation}
Here, $\epsilon_1$ and $\epsilon_2$ denote the energy shifts arising from the coupling for the basis states $|nlk\rangle$ and $|(n+1)l(k+1)\rangle$, respectively, while $\tilde{G}$ represents the effective coupling strength. These are the three coefficients to be determined in this ansatz. It is worth noting that we have omitted the common unperturbed energy contribution shared by both basis states, given by $n\Delta_{a}+l\tilde{\omega}_{b}+k\Delta_{c}$.
\begin{figure}[t]
	\centering
	\includegraphics[width=0.96\linewidth]{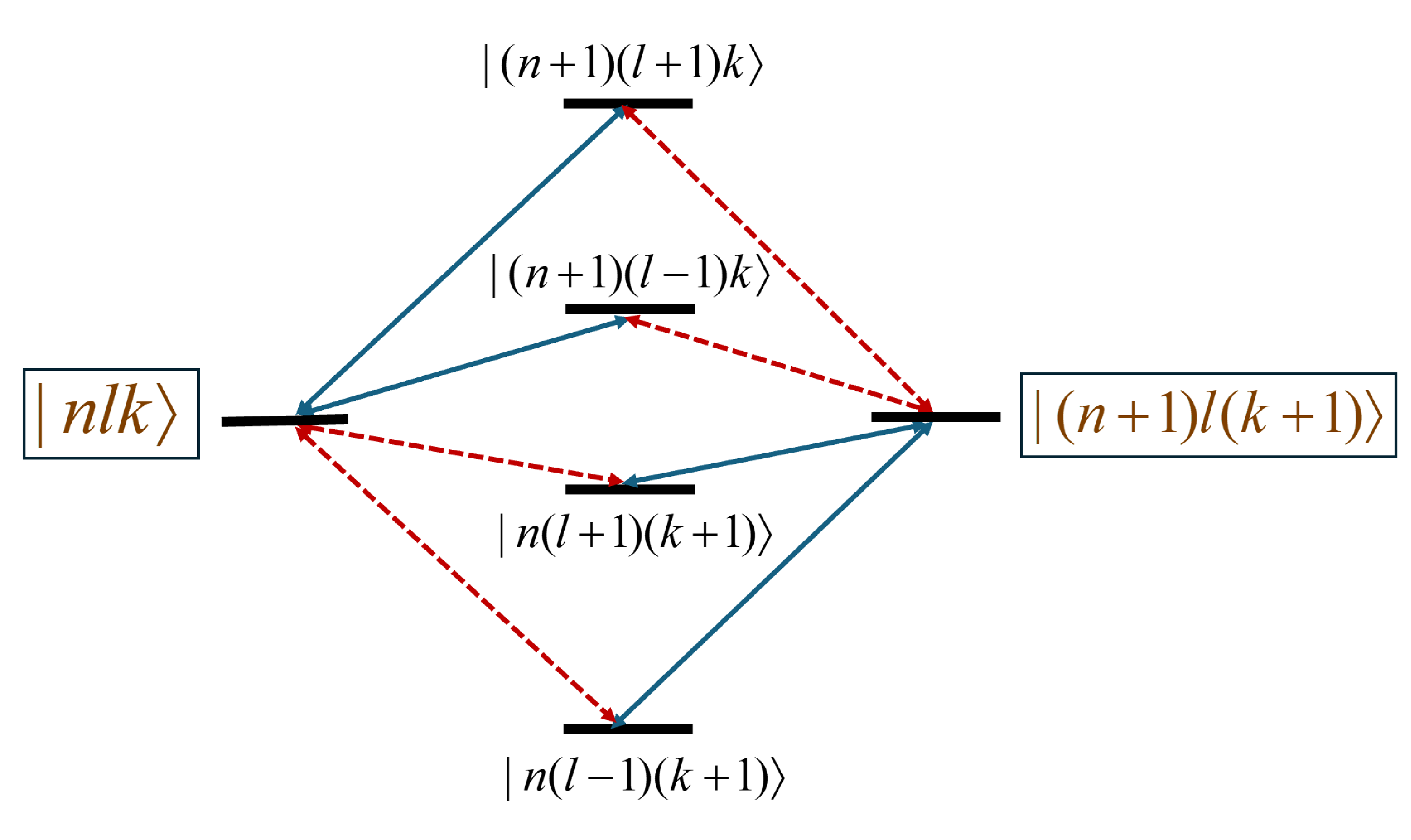}
	\caption{All evolution paths between the quantum states $|nlk\rangle$ and $|(n+1)l(k+1)\rangle$ are shown, with blue solid lines representing connections established by the optical-mechanical interaction and red dashed lines indicating transitions mediated by the mechanical-microwave interaction.}
	\label{transition}
\end{figure}

We first derive the effective coupling strength $\tilde{G}$ accounting for the contributions from all four virtual transition paths connecting the states $|nlk\rangle$ and $|(n+1)l(k+1)\rangle$, as illustrated in Fig.~\ref{transition}. By virtue of Eq.~\eqref{effcoustrequ}, one can have
\begin{equation}
\begin{aligned}
	\tilde{G}&=\frac{2\sqrt{(n+1)(k+1)}\tilde{\omega}_{b}gGe^{2r}e^{i(\alpha+\varphi)}}{\Delta_c^2-\tilde{\omega}_b^2}\\
	&\equiv\sqrt{(n+1)(k+1)}e^{i(\alpha+\varphi)}g_{\rm eff},
\end{aligned}
\end{equation}
up to the second order of $g$ and $G$.

In the following, we evaluate the energy shift $\epsilon_1$ associated with the state $|nlk\rangle$. Summarizing all eight paths from $|nlk\rangle$ to $|nlk\rangle$ through an intermediate state, four representative paths are presented in Fig.~\ref{transition}, one can obtain the second-order energy correction $\epsilon_1$ according to Eq.~\eqref{effcoustrequ}
\begin{equation}
\begin{aligned}
	\epsilon_1&=\frac{(n-l)G^2e^{2r}}{\Delta_a-\tilde{\omega}_{b}}-\frac{(n+l+1)G^2e^{2r}}{\Delta_a+\tilde{\omega}_b}\\
	&+\frac{(l-k)g^2e^{2r}}{\tilde{\omega}_b-\Delta_{c}}-\frac{(l+k+1)g^2e^{2r}}{\tilde{\omega}_b+\Delta_c}.
\end{aligned}
\end{equation}
Similarly, the energy shift of the state $|(n+1)l(k+1)\rangle$ can be derived as
\begin{equation}
\begin{aligned}
	\epsilon_2&=\frac{(n-l+1)G^2e^{2r}}{\Delta_a-\tilde{\omega}_{b}}-\frac{(n+l+2)G^2e^{2r}}{\Delta_a+\tilde{\omega}_b}\\
	&+\frac{(l-k-1)g^2e^{2r}}{\tilde{\omega}_b-\Delta_{c}}-\frac{(l+k+2)g^2e^{2r}}{\tilde{\omega}_b+\Delta_c}.
\end{aligned}
\end{equation}

To achieve exact resonance between the arbitrary states $|nlk\rangle$ and $|(n+1)l(k+1)\rangle$, the first two terms in Eq.~\eqref{effHamequ} must act as the identity operator within the relevant subspace. Thus, $\Delta_a+\Delta_c+\epsilon_2=\epsilon_1$. Assuming the detuning difference between $\Delta_a$ and $\Delta_c$ is given by $\delta$, we can then have
\begin{equation}
\delta\equiv\Delta_a+\Delta_c=\epsilon_1-\epsilon_2=\frac{2(G^2+g^2)e^{2r}\tilde{\omega}_{b}}{\tilde{\omega}_{b}^2-\Delta_c^2},
\end{equation}
up to the second order of the coupling strengths $g$ or $G$. Note $\delta$ is a Fock-state-independent coefficient. Under this condition, the effective Hamiltonian in Eq.~\eqref{effHamequ} can be simplified as
\begin{equation}\label{effHamequation}
\begin{aligned}
	H_{\text{eff}}&=\tilde{G}|nlk\rangle \langle (n+1)(k+1)|+\text{H.c.}\\
	&=\left[\tilde{G}|nk\rangle\langle(n+1)(k+1)|+\text{H.c.}\right]\otimes |l\rangle_b\langle l|.
\end{aligned}
\end{equation}

After eliminating the decoupled mechanical mode and extending the effective Hamiltonian in Eq.~\eqref{effHamequation} to the full Hilbert space of the optical and microwave modes, the resulting effective Hamiltonian eventually reduces to
\begin{equation}\label{Heffapp}
\begin{aligned}
	H_{\text{eff}}=g_{\text{eff}}[e^{i(\alpha+\varphi)}a^\dag c^\dag+e^{-i(\alpha+\varphi)}ac],
\end{aligned}
\end{equation}
where
\begin{equation}\label{geffapp}
\begin{aligned}
	g_{\text{eff}}=\frac{2\tilde{\omega}_{b}gGe^{2r}}{\Delta_c^2- \tilde{\omega}_b^2}.
\end{aligned}
\end{equation}
That is exactly the effective Hamiltonian given in Eq.~\eqref{effHam}.

\section{Validity of the effective Hamiltonian}\label{validity}
\begin{figure}[b]
\centering
\includegraphics[width=0.96\linewidth]{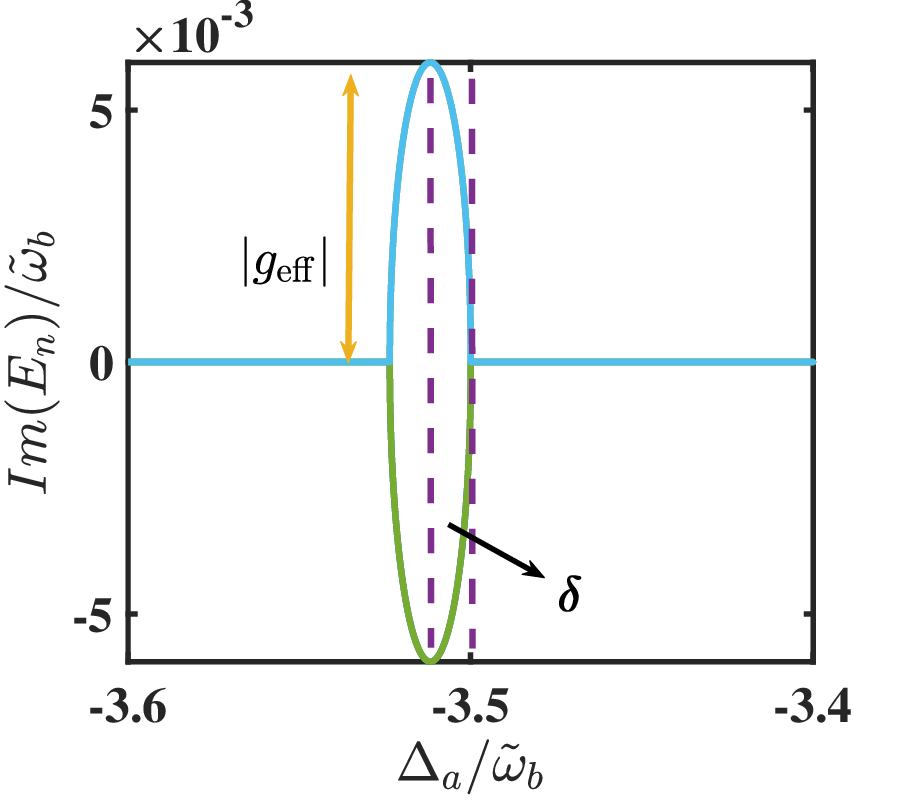}
\caption{Two relevant imaginary components of the normalized eigenvalues of the transition matrix are plotted versus the detuning frequency $\Delta_a$. The parameters used are $\Delta_c=3.5\tilde{\omega}_b$, $g=G=0.15\tilde{\omega}_b$, and $r=0.2$.}
\label{Liouvillian}
\end{figure}

The effective Hamiltonian presented in Eq.~\eqref{effHam} can be confirmed by diagonalizing the whole system transition matrix~\cite{photon-phononsqueezing}. Using the Heisenberg equations, the time-dependent quadrature operators governed by the full Hamiltonian in Eq.~\eqref{effHam} satisfy
\begin{equation}\label{Heisenequ}
\begin{aligned}
	\dot{\tilde{u}}(t)=i[H,\tilde{u}(t)]=i\mathcal{L}\tilde{u}(t),
\end{aligned}
\end{equation}
where $\tilde{u}(t)=[X_a(t),Y_a(t),X_b(t),Y_b(t),X_c(t),Y_c(t)]^T$
is the vector of quadrature operators, and $X_a=(e^{-i\alpha}a+e^{i\alpha}a^\dag)/\sqrt{2}$, $Y_a=(e^{-i\alpha}a-e^{i\alpha}a^\dag)/i\sqrt{2}$, $X_b=(b+b^\dag)/\sqrt{2}$, $Y_b =(b-b^\dag)/i\sqrt{2}$, $X_c=(e^{-i\varphi}c+e^{i\varphi}c^\dag) /\sqrt{2}$, $Y_c=(e^{-i\varphi}c-e^{i\varphi}c^\dag)/i\sqrt{2}$. 
$\mathcal{L}$ represents the system's transition matrix,
\begin{equation}\label{entireLiou}
\begin{aligned}
\mathcal{L}=-i\begin{bmatrix}
	0 & \Delta_a & 0 & 0 & 0 & 0 \\
	-\Delta_a & 0 & -2Ge^r & 0 & 0 & 0 \\
	0 & 0 & 0 & \tilde{\omega}_b & 0 & 0 \\
	-2Ge^r & 0 & -\tilde{\omega}_b & 0 & -2ge^r & 0\\
	0 & 0& 0& 0 & 0 & \Delta_{c} \\
	0 & 0 & -2ge^r & 0 & -\Delta_{c} & 0
\end{bmatrix}.
\end{aligned}
\end{equation}
As shown in the energy-level diagram of this matrix $\mathcal{L}$ versus $\Delta_a$, the presence of the two-mode squeezing interaction described by Eq.~\eqref{effHam} is manifested by the maximal splittings appearing in the imaginary parts of the eigenvalues~\cite{photon-phononsqueezing}.

\begin{figure}[t]
\centering
\includegraphics[width=1\linewidth]{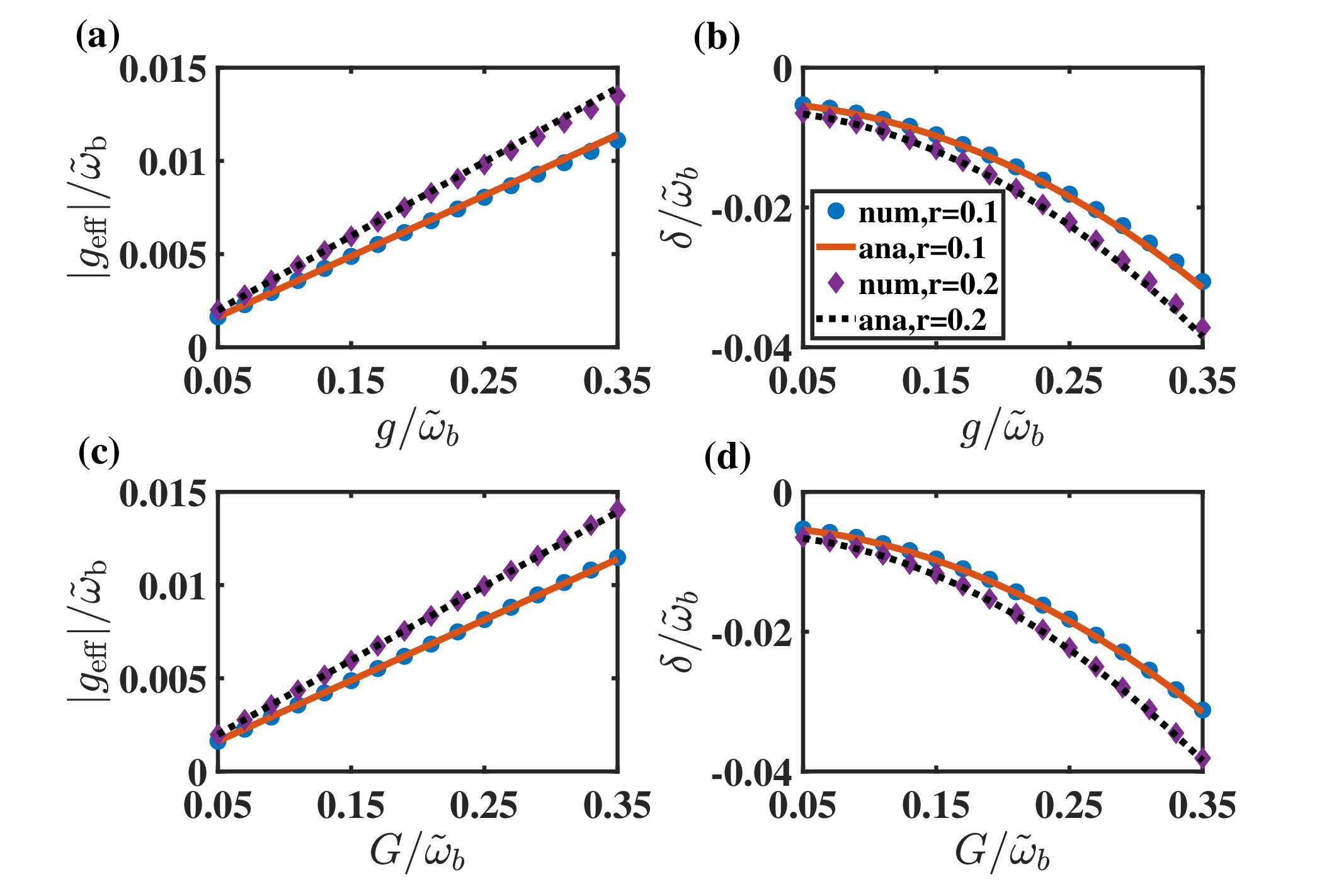}
\caption{[(a), (c)] Theoretical results about the effective coupling strength $g_{\rm eff}$ from Eq.~\eqref{effcoustr} (lines) are plotted as functions of $g$ and $G$, respectively, and compared against the numerical results, represented by discrete points. [(b), (d)] Analytical results for the energy shift $\delta$ in Eq.~\eqref{effcoustr} (lines) are similarly compared with the numerically obtained values (points) as functions of $g$ and $G$, respectively. $G=0.15\tilde{\omega}_b$ for panels (a) and (b), and $g=0.15\tilde{\omega}_b$ for panels (c) and (d). Additionally, $\Delta_c=3.5\tilde{\omega}_b$ is fixed for all panels.}
\label{numandana}
\end{figure}

In Fig.~\ref{Liouvillian}, we show the imaginary parts of the two relevant eigenvalues of the full transition matrix $\mathcal{L}$ in Eq.~\eqref{entireLiou}, obtained via numerical diagonalization. As the detuning of the optical mode $\Delta_a$ approaches (but does not exactly equal) the opposite detuning of the microwave mode $-\Delta_c$, a clear energy splitting emerges between the two eigenstates. The maximal splitting corresponds to the magnitude of $|g_{\rm eff}|$, while the energy shift $\delta$ arises from the mutual interaction between the microwave (optical) and mechanical modes.

The maximal splitting $|g_{\rm eff}|$ of the imaginary parts of the two eigenvalues [see Fig.~\ref{Liouvillian}] is presented in Figs.~\ref{numandana}(a) and~\ref{numandana}(c) as a function of the original coupling strengths $g$ and $G$, respectively. Blue dots and purple diamonds represent the numerical results for parameters $r=0.1$ and $r=0.2$, respectively, while the red solid and black dotted lines indicate the corresponding analytical predictions. In Fig.~\ref{numandana}(a), the effective coupling strength $g_{\rm eff}$ agrees well with their numerical results for $g\le 0.3\tilde{\omega}_b$ at $r=0.1$. As $r$ increases to $0.2$, the valid range reduces to $g/\tilde{\omega}_b\le0.25$. In Fig.~\ref{numandana}(c), $g_{\rm eff}$ remains valid for $G/\tilde{\omega}_b\le0.35$, independent of $r$. Similarly, the energy shift $\delta$ in Eq.~\eqref{effcoustr} is validated in Figs.~\ref{numandana}(b) and~\ref{numandana}(d), showing only slight deviations from numerical results when $g, G\ge0.3\tilde{\omega}_b$, with the derivation increasing gradually as $r$ increases.

\section{Optical-microwave squeezing in Markovian noises}\label{squeemarkovian}
In this section, we adopt an open-quantum-system framework to further evaluate the effective Hamiltonian given in Eq.~\eqref{effHam}, with a particular focus on its capacity to capture the generation of optical-microwave squeezing. In the presence of Markovian environments, the system's dynamics are governed by the quantum Langevin equation (QLE), which can be formulated in matrix form as
\begin{equation}\label{QLE}
\begin{aligned}
	\dot{u}(t)=Au(t)+n(t),
\end{aligned}
\end{equation}
The transition matrix is
\begin{equation}
\begin{aligned}
A=-\begin{bmatrix}
	\kappa_a & 0 & 0 & g_{\rm{eff}} \\
	0 & \kappa_a & g_{\rm{eff}} & 0 \\
	0 & g_{\rm{eff}} & \kappa_c & 0 \\
	g_{\rm{eff}} & 0 & 0 & \kappa_c
\end{bmatrix},
\end{aligned}
\end{equation}
where $\kappa_a$ and $\kappa_c$ represent the decay rates of modes $a$ and $c$, respectively. $u(t)=[X_a(t),Y_a(t),X_c(t),Y_c(t)]^T$ is the vector of quadrature operators. $n(t)=[X^{in}_a(t),Y^{in}_a(t),X^{in}_c(t),Y^{in}_c(t)]^T$ is the vector of Gaussian noise operators, and $X^{in}_o=(e^{-i\alpha}a_{in}+e^{i\alpha}a^\dag_{in})/{\sqrt{2}}$, $Y^{in}_a=(e^{-i\alpha}a_{in}-e^{i\alpha}a^\dag_{in})/{i\sqrt{2}}$, $X^{in}_c=(e^{-i\varphi}c_{in}+e^{i\varphi}c^\dag_{in})/{\sqrt{2}}$, $Y^{in}_c=(e^{-i\varphi}c_{in}-e^{i\varphi}c^\dag_{in})/{i\sqrt{2}}$. $a_{in}$ and $c_{in}$ are characterized by their respective covariance functions, $\langle o^\dag_{in}(t)o_{in}(t')\rangle=N_o\delta(t-t'), ~o=a,c$, in which $N_o=[\exp(\hbar\omega_o/k_BT)-1]^{-1}$ is the mean population of mode $o$ at the thermal equilibrium state.

\begin{figure}[t]
	\centering
	\includegraphics[width=0.96\linewidth]{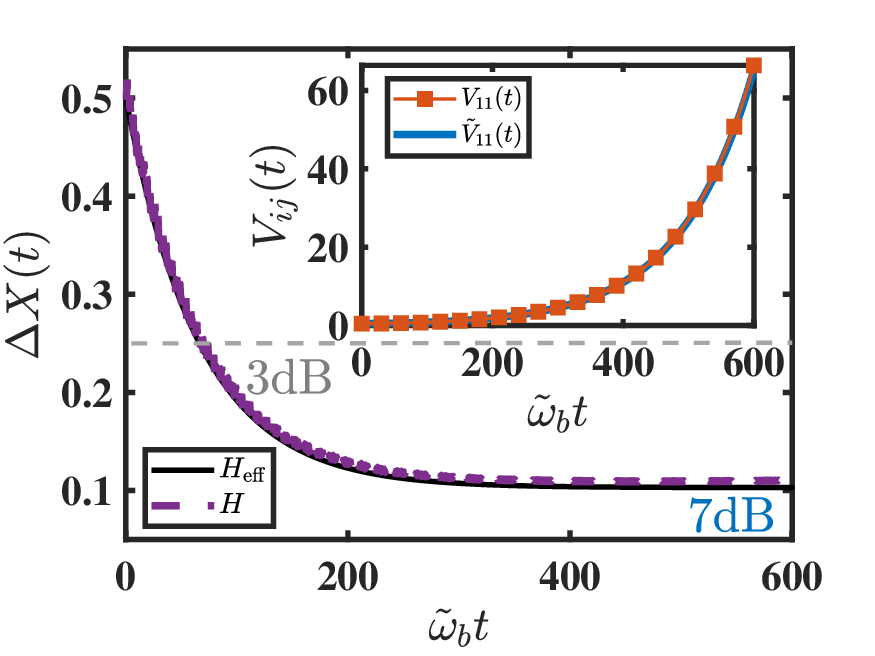}
	\caption{The dynamics of the variance $\Delta X(t)$ are presented using the effective Hamiltonian in Eq.~\eqref{effHam} and the full system Hamiltonian in Eq.~\eqref{linHam}. The inset depicts the evolution of the corresponding CM element under the effective and full Hamiltonians. The parameters are set as $\kappa_b=10^{-5}\tilde{\omega}_b$, $\kappa_a=0.002\tilde{\omega}_b$, $\kappa_c=10^{-3}\tilde{\omega}_b$, and the thermal numbers $N_a=N_c=0$ and $N_b=10$. Other parameters are the same as Fig.~\ref{Liouvillian}.}
	\label{CMandvariance}
\end{figure}

According to the QLE in Eq.~\eqref{QLE}, the system's evolution can be completely characterized by a $4\times4$ CM $V(t)$. The dynamics of the CM $V(t)$ are governed by
\begin{equation}\label{CM}
\begin{aligned}
	\dot{V}(t)=AV(t)+V(t)A^T+D,
\end{aligned}
\end{equation}
The entries of $V(t)$ are defined as
\begin{equation}
\begin{aligned}
	V_{ij}(t)=\frac{\langle u_i(t)u_j(t)+u_j(t)u_i(t)\rangle}{2},
\end{aligned}
\end{equation}
where $u_i(t)$ is the $i$-term of $u(t)$ and $i=1,2,3,4$. The diffusion matrix $D=\text{Diag}[\kappa_a(2N_a+1),\kappa_a(2N_a+1),\kappa_c(2N_c+1), \kappa_c(2N_c+1)]$, where the elements are determined by $ D_{ij}(t)={\langle n_i(t)n_j(t)+n_j(t)n_i(t)\rangle}/{2}$. Under steady-state conditions, the CM remains invariant over time, meaning $\dot{V}=0$, which necessitates $g_{\rm eff}^2<\kappa_a\kappa_c$.

In the unsteady-state regime, $g^2_{\rm eff}>\kappa_a\kappa_c$, all the CM elements exhibit exponential divergence. Under this condition, there exists an optimized two-mode squeezing operator
\begin{equation}\label{squeeoperator}
	X=\cos\phi X_a+\sin\phi Y_c,\quad \tan(2\phi)=\frac{2g_{\text{eff}}}{\kappa_a-\kappa_c}
\end{equation}
to realize the stable optical-microwave squeezing. By virtue of Eq.~\eqref{CM}, the variance $\Delta X=\langle X^2\rangle-\langle X\rangle^2$ can be analytically derived as
\begin{equation}\label{deltaxvariance}
	\Delta X(t)=\frac{1}{2}+2Ce^{-(\Omega+\kappa_a+\kappa_c)t}-2C,
\end{equation}
where 
\begin{equation}\label{CMparameters}
\begin{aligned}
	\Omega&=\sqrt{4g^2_{\text{eff}}+(\kappa_a-\kappa_c)^2},C=-\frac{\kappa_++\cos(2\phi)\kappa_-}{4(\Omega+\kappa_a+\kappa_c)}+\frac{1}{4},\\
	\kappa_\pm&=\kappa_a(2N_a+1)\pm\kappa_c(2N_c+1).
\end{aligned}
\end{equation}
Here, we assume the initial CM is $V(0)={I_4}/{2}$, where $I_4$ is the identity matrix of four dimensions.

The results obtained by the effective Hamiltonian in Eq.~\eqref{effHam} can be further validated by examining the full system dynamics. Analogous to Eq.~\eqref{CM}, the time evolution of the CM $\tilde{V}(t)$ of the whole system, based on the full system Hamiltonian $H$ in Eq.~\eqref{linHam}, is governed by
\begin{equation}\label{entireCM}
\begin{aligned}
	\dot{\tilde{V}}(t)=\tilde{A}\tilde{V}(t)+\tilde{V}(t)\tilde{A}^T+\tilde{D}.
\end{aligned}
\end{equation}
The elements of $\tilde{V}(t)$ are defined as
\begin{equation}
	\tilde{V}_{ij}(t)=\frac{\langle\tilde{u}_i(t)\tilde{u}_j(t)+\tilde{u}_j(t)\tilde{u}_i(t)\rangle}{2}, i,j=1,2,\dots6,
\end{equation}
where $\tilde{u}(t)$ is shown in Eq.~\eqref{Heisenequ}. The transition matrix $\tilde{A}=i\mathcal{L}-K$, where $\mathcal{L}$ is the matrix in Eq.~\eqref{entireLiou} and $K=\text{Diag}[\kappa_a, \kappa_a,e^{2r}\kappa_b,e^{2r}\kappa_b,\kappa_c,\kappa_c]$. $\tilde{D}=\text{Diag}[\kappa_a(2N_a + 1),\kappa_a(2N_a + 1),e^{2r}\kappa_b(2N_b + 1),e^{2r}\kappa_b(2N_b + 1),\kappa_c(2N_c + 1),\kappa_c(2N_c + 1)]$ is the matrix of noise covariance. The mechanical decay rate is exponentially enlarged due to the mechanical amplifier effect, i.e., $\kappa_b\to e^{2r}\kappa_b$\cite{Optomechanicalsqueezing}. Then, the dynamics of $\Delta X(t)$ can be achieved by numerically calculating the CM $\tilde{V}(t)$,
\begin{equation}
	\Delta X(t)
	=\cos^2\phi\tilde{V}_{11}(t)+\sin^2\phi\tilde{V}_{44}(t)+\sin(2\phi)\tilde{V}_{14}(t),
\end{equation}
Here, the initial condition is $\tilde{V}(0)=I_6/2$, and $I_6$ is a six-dimensional identity matrix. 

Numerical results for the variance $\Delta X(t)$ are shown in Fig.~\ref{CMandvariance}. It can be observed that the values of $\Delta X(t)$ (purple dashed line) obtained using Eq.~\eqref{entireCM} do match well with those calculated via the effective Hamiltonian [Eq.~\eqref{deltaxvariance}, black solid line]. Additionally, the matrix element $\tilde{V}_{11}(t)$, calculated by Eq.~\eqref{entireCM}, is presented in the inset and also exhibits good agreement with the corresponding effective result $V_{11}(t)$ (red-solid line with square markers).

\section{Variance derivation under non-Markovian noises}\label{appderivation}
This appendix presents the detailed derivation of the variance $\Delta X(t)$ in the presence of non-Markovian environmental noises. Using Eq.~\eqref{nmhl}, the Green's functions $\mathcal{U}(t)$ and $\mathcal{V}(t)$ introduced in the Sec.~\ref{ppsqueeze} satisfy the following Dyson equations,
\begin{equation}\label{b1}
\begin{aligned}
	\dot{\mathcal{U}}(t)&=T\mathcal{U}(t)-\int_0^t \bar{F}(t-s)\mathcal{U}(s)\mathrm{d}s, \\
	\dot{\mathcal{V}}(t)&=T\mathcal{V}(t)-\int_0^t \bar{F}(t-s)\mathcal{V}(s)\mathrm{d}s+\epsilon_{\text{in}}(t).
\end{aligned}
\end{equation}
Considering the initial conditions $\mathcal{U}(0)=I$ and $\mathcal{V}(0)=0$, one can have
\begin{equation}\label{vtnon}
\begin{aligned}
\mathcal{V}_1(t)&\!=\!\int_0^t\mathrm{d}s\bigl[\mathcal{U}_{11}(t-s) a_{in}(s)\!+\!\mathcal{U}_{12}(t-s)c_{in}^\dagger(s)\bigr], \\
\mathcal{V}_2(t)&\!=\!\int_0^t\mathrm{d}s\bigl[\mathcal{U}_{21}(t-s) a_{in}(s)\!+\!\mathcal{U}_{22}(t-s)c_{in}^\dagger(s)\bigr].	
\end{aligned}
\end{equation}	

Owing to the definitions of Gaussian noises $a_{in}$ and $c_{in}$ provided in Sec.~\ref{ppsqueeze}, it is possible to further derive the auto-correlation term
\begin{equation}
\begin{aligned}
	\langle\mathcal{V}_1\mathcal{V}^\dag_1\rangle&=\int \mathrm{d}\omega J_a(\omega)\left[n_a(\omega)+1\right]|\tilde{U}_{11}(t)|^2\\
	&\quad+\int \mathrm{d}\omega J_c(\omega)n_c(\omega)|\tilde{U}_{12}(t)|^2,\\ 
\end{aligned}
\end{equation}
and the cross-correlation term
\begin{equation}
\begin{aligned}
	\langle\mathcal{V}_1\mathcal{V}^\dag_2\rangle&=\int \mathrm{d}\omega J_a(\omega)[n_a(\omega)+1]\tilde{U}_{11}(t)\tilde{U}^*_{21}(t)\\
	&\quad+\int \mathrm{d}\omega J_c(\omega)n_c(\omega)\tilde{U}_{12}(t)\tilde{U}^*_{22}(t),\\
\end{aligned}
\end{equation}
where 
\begin{equation}
	\tilde{U}_{kj}(t)=\int^t_0 \mathcal{U}_{kj}(t-s)e^{i(-1)^j\omega s}\mathrm{d}s,~k,j=1,2\\
\end{equation}
and $U^*_{kj}(t)$ is the complex conjugate of $U_{kj}(t)$. $n_a(\omega)$ and $n_c(\omega)$ are the average numbers at thermal states, respectively. The other terms about $\langle\mathcal{V}^\dag_1\mathcal{V}_2\rangle$,
$\langle\mathcal{V}^\dag_2\mathcal{V}_1\rangle$,
$\langle\mathcal{V}_2\mathcal{V}^\dag_2\rangle$, $\langle\mathcal{V}^\dag_2\mathcal{V}_2\rangle$, $\langle\mathcal{V}_2\mathcal{V}^\dag_1\rangle$, and $\langle\mathcal{V}^\dag_1\mathcal{V}_2\rangle$ can be also similarly derived.

With the solutions about $\mathcal{U}(t)$ and $\mathcal{V}(t)$, the dynamical evolution of $a(t)$ and $c^\dag(t)$ shown in Eq.~\eqref{nmhl} can be written as
\begin{equation}\label{abtu}
\begin{aligned}
	a(t)&=\mathcal{U}_{11}(t)a(0)+\mathcal{U}_{12}(t)c^\dagger(0)+\mathcal{V}_1(t), \\
	c^\dagger(t)&=\mathcal{U}_{21}(t)a(0)+\mathcal{U}_{22}(t)c^\dagger(0)+\mathcal{V}_2(t).
\end{aligned}
\end{equation}	
The system's dynamics can be completely characterized by a $4\times 4$ CM $V(t)$. Using the CM definition given in Eq.~\eqref{CM}, one can obtain
\begin{equation}\label{CMelement}
\begin{aligned}
V_{11}&\!=\!\frac{|\mathcal{U}_{11}^2|\!+\!|\mathcal{U}_{12}^2|\!+\!\langle \mathcal{V}_1\mathcal{V}^\dag_1\rangle\!+\!\langle \mathcal{V}^\dag_1\mathcal{V}_1\rangle}{2},\\
V_{44}&\!=\!\frac{|\mathcal{U}_{21}^2|\!+\!|\mathcal{U}_{22}^2|\!+\!\langle \mathcal{V}_2\mathcal{V}^\dag_2\rangle\!+\!\langle \mathcal{V}^\dag_2\mathcal{V}_2\rangle}{2},\\
V_{14}&\!=\!\frac{e^{-i\tilde{\varphi}}(\mathcal{U}_{11}\mathcal{U}^*_{21}\!+\!\mathcal{U}^*_{22}\mathcal{U}_{12}\!+\!\langle\mathcal{V}_1\mathcal{V}^\dag_2\rangle\!+\!\langle\mathcal{V}^\dag_2\mathcal{V}_1\rangle)}{4i}\\
&\quad\!-\!\frac{e^{i\tilde{\varphi}}(\mathcal{U}^*_{12}\mathcal{U}_{22}\!+\!\mathcal{U}^*_{11}\mathcal{U}_{21}\!+\!\langle\mathcal{V}^\dag_1\mathcal{V}_2\rangle\!+\!\langle\mathcal{V}_2\mathcal{V}^\dag_1\rangle)}{4i},
\end{aligned}
\end{equation}
where $\tilde{\varphi}=\alpha+\varphi$, and other non-zeros elements are $V_{22}=V_{11}$, $V_{33}=V_{44}$, and $V_{23}=V_{14}$. Using the definitions of the squeezing operator $X$ in Eq.~\eqref{MOSX} and antisqueezing operator $Y$ in Eq.~\eqref{MOSX}, one can eventually obtain the variances $\Delta X(t)$ and $\Delta Y(t)$, shown in Eq.~\eqref{variXY}.

\bigskip

\bibliographystyle{apsrevlong}
\bibliography{reference}

\end{document}